\documentclass[journal]{IEEEtran}
\usepackage{cite}
\usepackage{array}
\usepackage{amsmath}
\usepackage{tabularx}
\usepackage{caption}
\usepackage{url}
\usepackage[latin1]{inputenc}
\usepackage[T1]{fontenc}
\usepackage{url}
\usepackage{booktabs}
\usepackage{amsmath}
\usepackage{relsize}
\usepackage{graphicx}
\usepackage{subfigure}
\usepackage{hyperref}
\usepackage{color}
\usepackage{complexity}

\begin{document}

\title{Using Genetic Algorithms to Build Adaptive Wireless Networks: A Survey}
\title{Genetic Algorithms in Wireless Networking: \\
\textit{Techniques, Applications, and Issues}}

\author{Usama Mehboob, Junaid Qadir, Salman Ali, and Athanasios Vasilakos

\thanks{Usama Mehboob (10beeumehboob@seecs.edu.pk), Junaid Qadir (junaid.qadir@seecs.edu.pk), and Salman Ali (09mscsesali@seecs.edu.pk) are with the Electrical Engineering Department at the School of Electrical Engineering and Computer Science (SEECS), at the National University of Sciences and Technology (NUST), Pakistan. Junaid Qadir is an Assistant Professor at SEECS, NUST. Athanasios Vasilakos (vasilako@ath.forthnet.gr) is a Professor of Computer Engineering at the University of Western Macedonia, Greece.}}

\maketitle

\begin{abstract}

In recent times, wireless access technology is becoming increasingly commonplace due to the ease of operation and installation of untethered wireless media. The design of wireless networking is challenging due to the highly dynamic environmental condition that makes parameter optimization a complex task. Due to the dynamic, and often unknown, operating conditions, modern wireless networking standards increasingly rely on machine learning and artificial intelligence algorithms. Genetic algorithms (GAs) provide a well-established framework for implementing artificial intelligence tasks such as classification, learning, and optimization. GAs are well-known for their remarkable generality and versatility, and have been applied in a wide variety of settings in wireless networks. In this paper, we provide a comprehensive survey of the applications of GAs in wireless networks. We provide both an exposition of common GA models and configuration and provide a broad ranging survey of GA techniques in wireless networks. We also point out open research issues and define potential future work. While various surveys on GAs exist in literature, our paper is the first paper, to the best of our knowledge, which focuses on their application in wireless networks.
\end{abstract}

\section{Introduction}
\label{sec:introduction}

The study of nature is a rich source of inspiration for researchers working in the area of artificial intelligence and machine learning. In particular, the human brain and the biological process of evolution have helped develop and guide research in the neural network, and the evolutionary algorithms communities.  A genetic algorithm (GA) is a metaheuristic computational method \cite{hillier2001introduction}, inspired from biological evolution \cite{ridley2004evolution}, that aims to imitate the robust procedures used by various biological organisms to adapt as part of their natural evolution. GAs have been successfully used in fields as diverse as aircraft industry, chip design, computer animation, drug design, telecommunications, software creation, and financial markets \cite{mitchell1998introduction}.

The field of genetic algorithms (GA) was established by John Holland who investigated the evolution of \emph{complex adaptive systems} (CAS) comprising interacting genes, starting in the early 1960s, and culminating in the publication in 1975 of his seminal book on this topic \cite{holland1975genetic}. The idea of embedding computers with learning inspired from evolution goes back even further as Alan Turing proposed a ``learning machine'' which would parallel the principles of evolution in his landmark 1950 paper on machine intelligence\cite{turing1950computing}. Holland used the biological metaphor of \textit{chromosomes} to refer to strings of binary symbols encoding a candidate solution to the given problem. Holland proposed using the computational analogues of the biological evolutionary processes of random \textit{mutation}, \textit{crossover}, and natural \textit{selection} \cite{goldberg1988genetic} to enable populations of chromosomes to get increasingly better at problem solving. The underlying premise is that given enough time, the process would converge towards a population that contains a chromosome (or chromosomes) that solves the given problem. Apart from the mutation and crossover operators, The design of a GA framework also involves other design issues such as genetic representation (encoding), population initialization, fitness function formulation, and a selection mechanism. More recently, researchers have also proposed the use of altruistic techniques in an evolutionary framework that involve cooperation as a fundamental primitive \cite{highfield2011supercooperators}. 

\subsection{Motivations for using GA} 

\vspace{2mm}
\subsubsection{Generality and Versatility} 

GAs apply in a wide variety of settings and can be easily molded to particular problems. GAs constitute a very general meta-heuristic technique which can be thought of as the sledgehammer of the craft of algorithms, much like the technique of artificial neural networks (ANNs). In particular, GAs can be readily invoked in areas that do not yield readily to standard approaches, or when more specialized techniques fail. GAs are capable of solving extremely large problems that have large search spaces. GAs are very good at navigating through huge search spaces to heuristically find near optimal solutions in quick time. In addition, GAs can work even when the objective function is not exactly known since GAs rely only on an objective function's evaluation (without necessarily knowing the objective function explicitly). Although GAs do not guarantee optimality, GAs are generally useful in practice. Bhandar et al. \cite{bhandari1996genetic} showed that although GAs do not guarantee convergence to an optimal solution, GAs avoid local optimas with a high probability through the use of mutation and crossover operators. 

\vspace{2mm}
\subsubsection{Adaptiveness and Online Problem Solving} 

To understand adaptivity in changing conditions, we will initially develop the idea of a `\textit{landscape}' both as a metaphor (in which we visualize ourselves as climbing a landscape in pursuit of the highest peak) and as a mathematical object (in which the value of the function maps to the elevation of the landscape). Building upon the metaphor of fitness landscapes, and the insight of GAs as stochastic search algorithms, two landscape models of relevance to optimization through GAs have been proposed by Page \cite{page2010diversity}. \emph{Rugged landscapes} are landscapes in which there are many peaks, valleys, and troughs (and not a single peak). Rugged landscapes assume that the fitness levels do not change; in evolutionary systems, however, the fitness function is also dependent on context, and on the behavior of other competitors. This can be captured by the metaphor of \emph{dancing landscapes}---which are adapted from rugged landscapes. When a landscape `dances', local peaks may change, making a solution that was earlier optimal no longer a peak. 

Traditional techniques from optimization theory assume static and well-known topologies and are not suited to dynamic environments. In many problems of practical interest, our focus is more on \textit{satisficing} (i.e., finding a sufficiently good solution that satisfies one's purposes) rather than on \textit{optimizing} (i.e., finding the best possible solution). A key benefit of GAs is that they are well suited to the optimization task in the dancing landscapes that characterize wireless networks (in which there is interaction between the various adaptive wireless nodes and interdependency between the various node's decisions). GAs are well suited for building metaheuristic adaptive algorithms that can provide satisfactory performance in changing network conditions. 

Another important aspect of GAs is that it is an \textit{online} adaptive algorithm that that can operate in unknown environments in an online fashion. Such an ability is crucial in various control settings in wireless networks in which decisions have to be made automatically in run-time to cater to dynamic channel parameters. In dynamic channel conditions that typically exist in most wireless networking configurations, optimization is an elusive moving target since the optimal solution keeps changing as the conditions change. In online adaptive optimization algorithms, there is a fundamental tradeoff between \textit{exploration}---which involves looking for potentially better previously unexplored solutions---and \textit{exploitation} that implies the use of previously known good solutions. Exploration in effect is an attempt to find good adaptive building blocks and exploitation is the use and propagation of adaptations that are known to perform well. We can envision mutation and recombination of genes as analogous to exploration, whereas natural selection can be envisioned as a form of exploitation \cite{page2010diversity}. In ever changing dynamic environments, such as wireless networks, it is important to always keep on exploring. Holland's original GA work used schema analysis to show that, under qualifying assumptions, GAs can achieve near-optimal exploration and exploitation tradeoff.

\vspace{2mm}
\subsubsection{Ability to Find Good Building Blocks}

By working in terms of a population of candidate solutions, genetic algorithms can exploit the diversity of solutions to find building blocks---known as schemas in literature---which are substrings of chromosomes that denote high performing elements of the overall solution \cite{holland1975genetic}. In the context of biological genetics, schemas correspond to constellations of genes working together to help an organism adapt; evolution proceeds by selecting these constellations  (or building blocks) through natural selection. Using genetic operators such as recombination and crossover, GAs can evolve better solutions through mixing of the high performing basic building blocks.

\vspace{2mm}
\subsubsection{Parallel Nature and Scalability} 

Traditional theory of GAs presumes that GAs accomplish the culmination by discovering, emphasizing, and recombining the \emph{good traits} of chromosomes in a vastly parallel manner. GAs incorporate multiple solutions together in a `population' of solutions. GAs use evolutionary techniques to test and improve the solutions by using techniques such as mutation, crossover, selection, and recombination. The parallel nature of GAs renders it as a suitable optimization framework for scalable problem solving in a wide range of applications.

\vspace{2mm}
\subsubsection{Support For Multiobjective Optimization}

Many practical problems in wireless networks require optimizing for multiple parameters (that may potentially conflict with each other). An important characteristic of GAs is that it can easily support joint optimization of multiple objectives. Support for multi-objective optimization---in which the considered objectives may potentially conflict with each other---is of significant practical interest in wireless networks. Issues related to multi-objective optimization are introduced in \cite{konak2006multi}.

\vspace{2mm}
\subsubsection{Support for Global Optimization} 

Unlike network models such as the multi-layered perceptrons (which make local changes so as to find the local minima), GAs are suited to finding the global optima due to a number of properties:

\begin{itemize}
\item They search by means of a population of individuals.
\item They work with an encoding of the multiple parameters.
\item They use a fitness function that does not require the calculation of derivatives.
\item They search probabilistically. 
\end{itemize}

\vspace{2mm}
\subsubsection{Easy Implementation} 

GAs are computationally simpler compared to other complementary AI techniques such as neural networks since they require only swapping and shifting of genes in chromosomes (unlike neural networks that require adders for its multiple hidden layers). GAs are easily implemented and can be rapidly prototyped in digital signal processors (DSPs) or field programmable gate arrays (FPGAs). In addition, GAs are highly amenable to successful implementation on modern parallel and cloud computing architectures due to its inherently parallel nature. 

\subsection{Contributions of this Paper}

This paper offers a comprehensive survey of the applications of GAs in wireless networking. In addition, we also provide a self-contained introduction to the GAs. The aim of this paper is not to provide an exhaustive survey\footnote{The sheer breadth of GA literature precludes an exhaustive survey, the exhaustive bibliography created by Goldberg et al. in 1997---which is a excellent reference to the pool of GA literature---amassed more than 4000 publications. With the unabated interest in GAs, it will be no surprise if an updated bibliography contains more than double the original references.}, but to provide a  representative sample of important works along with a comprehensive listing of the various applications of GAs. We also highlight the application of GAs in wireless networking configurations such as wireless mesh networks (WMNs), mobile ad-hoc networks (MANETs), cellular networks (CNs), wireless sensor networks (WSNs), and cognitive radio networks (CRNs). While a lot of material exists on GAs---including various textbooks \cite{goldberg1988genetic} \cite{davis1991handbook} \cite{pedrycz2010computational}, as well as generic survey papers \cite{srinivas1994genetic} \cite{man1996genetic}---a focused survey article on the applications of GAs in wireless networking is missing in literature. This paper will be useful to networking researchers interested in applying GAs in particular, and metaheuristic and evolutionary techniques more generally, in the context of wireless networks. According to the best of our knowledge, this is the first survey paper that provides an extensive survey of the applications of GAs in general wireless networks. 

\subsection{Organization of this Paper}

We provide the necessary background on GA theory, working and common techniques used to carry out the evolutionary process in Section \ref{sec:background}. The applications of GAs in wireless networks---such as routing, channel allocation, load balancing, localization, and quality of service (QoS)---are listed in  Section \ref{sec:applications}. We highlight the various lessons learnt through years of GA research, in terms of common pitfalls as well as guidelines for successful implementation, in Section \ref{sec:insights}. In Section \ref{sec:openissues}, we highlight open issues and suggest directions for future work. Finally, this paper is concluded in Section \ref{sec:conclusions}.

\vspace{3mm}
To facilitate the reader, acronyms used in this paper are collected in Table \ref{tab:acronyms} as a convenient reference.

\begin{table}
\caption{Acronyms used in this paper.}
\label{tab:acronyms}
\footnotesize
\centering
\begin{tabular}{p{1.3cm}p{5.3cm}}

\toprule
\textbf{\textit{Acronym}} & \textbf{\textit{Expanded Form}}\\
\midrule

ACO  &  Ant Colony Optimization\\
AP  &  Access Point\\
ANN   &   Artificial Neural Network \\
BER  &  Bit Error Rate\\
BFWA  &  Broadband Fixed Wireless Access\\
BS  &  Base Station\\
BSP  &  Broadcast Scheduling Problem\\
CAC  &  Call Admission Control\\
CBDT  &  Case Base Decision Theory\\
CSM  &  Cognitive System Module\\
CN  &  Cellular Network \\
CDMA   &   Code Division Multiple Access\\
CR   &   Cognitive Radio \\
CRN   &   Cognitive Radio Network \\
DCA  &  Dynamic Channel Assignment\\
DSA & Dynamic Spectrum Access\\
DSPRP  &  Dynamic Shortest Path Routing Problem\\
FA  &  Forward Ant\\
FCA  &  Fixed Channel Assignment\\
FPGA  &  Field Programmable Gate Array\\
GA  &  Genetic Algorithm\\
GP  &  Genetic Programming\\
GPRS  &  General Packet Radio Service\\
iGA  &  Island GA\\
HLB  &  Hybrid Load Balancing\\
ILP  &  Integer Linear Programming\\
LISP  &  List Processing (Programming Language)\\
LLF  &  Least Loaded First\\
LTE   &   Long Term Evolution\\
MAC   &   Media Access Control\\
MANET   &   Mobile Ad-hoc Network\\
MDP  &  Markov Decision Process\\
MOGA  &  Multi-Objective GA\\
MSC  &  Mobile Switching Center\\
NN  &  Neural Network\\
NSGA  &  Non-Dominated Sorted Genetic Algorithm\\
NSGA-II  &  Non-Dominated Sorted Genetic Algorithm-II\\
NSGS-IIa  &  Adaptive NSGA-II\\
NSGA-IIr  &  Random NSGA-II\\
PGA & Parallel GA\\
pNSGA-II  &  Parallel NSGA-II\\
QoS  &  Quality of Service\\
RNC  &  Radio Network Controller \\
RSS   &   Radio Signal Strength\\
RSSI  &  Received Signal Strength Indicator\\
SA  &  Simulated Annealing\\
SGSN  &  Serving GPRS Node\\
SINR  &  Signal to Interference and Noise Ratio\\
SNR   &   Signal to Noise Ratio\\
SOGA  &  Single Objective GA\\
SP  &  Shortest Path\\
ssNSGA-II  &  Steady State NSGA-II\\
SWN &  Software Defined Wireless Network\\
UMTS   &   Universal Mobile Telecommunications System\\
WLAN  &  Wireless Local Area Network\\
WMN   &   Wireless Mesh Network\\
WSGA  &  Wireless System Genetic Algorithm\\

\bottomrule
\end{tabular}
\end{table}

\section{Background: Genetic Algorithms}
\label{sec:background}

\begin{figure*}[ht!]
\centering
\includegraphics[width=.93\textwidth]{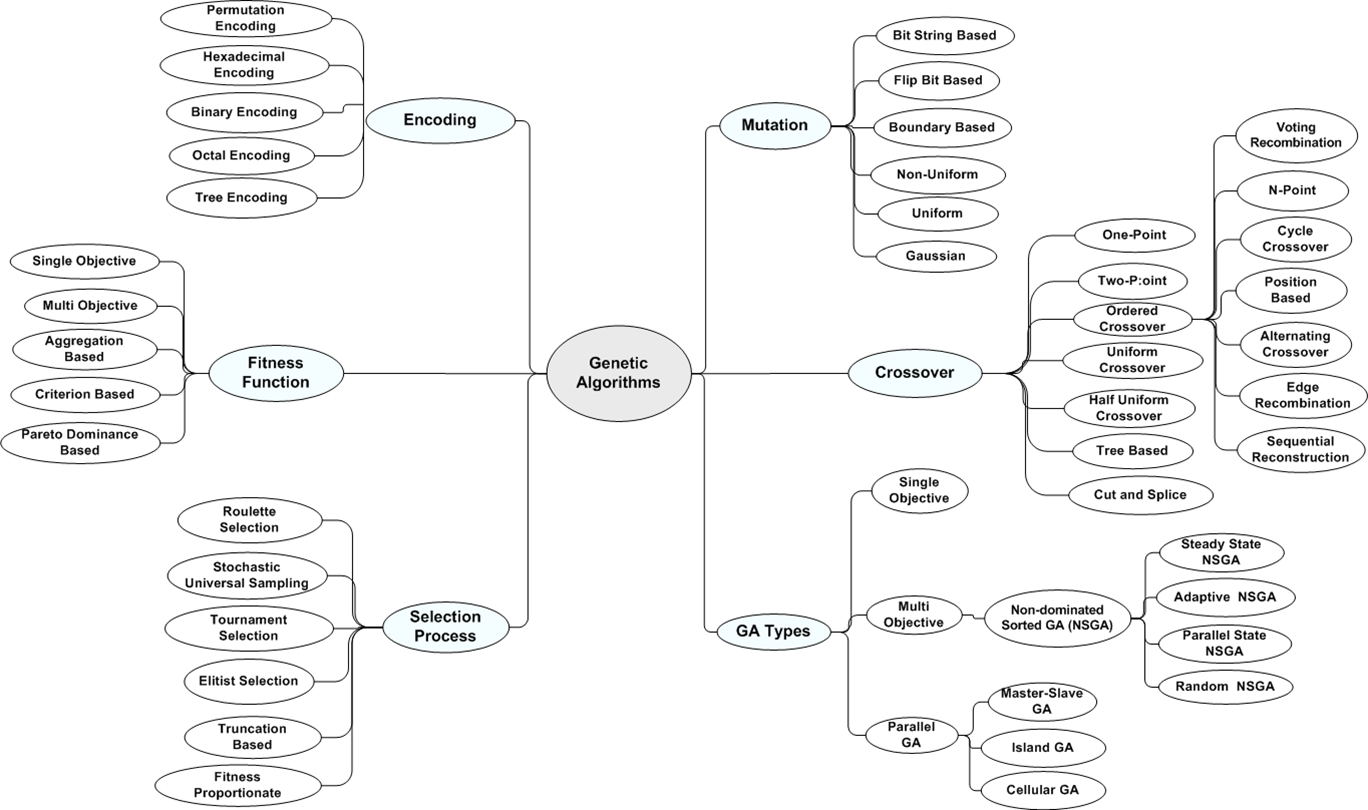}
\captionsetup{justification=centering,margin=2cm}
\caption{Taxonomy of Genetic Algorithms}
\label{fig:taxonomy}
\end{figure*}

A detailed GA taxonomy is provided in Figure \ref{fig:taxonomy}. This taxonomy diagram exhibits the breadth of GA's applications as described in this section.

\subsection{GA Terminology}

Since the field of GA is inspired from the biological genetic evolution, the field uses a number of biological metaphors in its terminology. The purpose of this short subsection is to introduce these biological metaphors \emph{in the context of applications of GAs in wireless networking}. We introduce the terminology through the following list. 

\begin{itemize}

\item \emph{Organism}: It represents the entity (e.g., a radio parameter, or a wireless resource) being optimized. 

\item \emph{Population}: It refers to the ensemble or collection of organisms undergoing simulated genetic evolution.

\item \emph{Chromosome}: It encodes a particular solution to the problem under study. (Biologically, a chromosome contains an organism's genetic makeup.)  

\item \emph{Fitness}: It represents the utility of the current chromosome. (In evolutionary theory, fit chromosomes are passed on through heredity while unfit chromosomes die out due to the natural phenomenon of the `survival of the fittest'.) 

\item \emph{Gene}: It is the basic building block of the chromosome defining a particular feature of the simulated organism. (Each chromosome can contain a 
number of genes.)

\item \emph{Allele}: Each gene can take several alternative forms, each of which is called an allele. 

\item \emph{Locus}: It is the position on the chromosome containing a particular gene of interest.

\item \emph{Mutation}: It represents a random change of an allele (or the value of a gene). 

\item \emph{Selection}: The process through which the `fittest' simulated organism survives while the unfit solutions are weeded out.

\end{itemize}

\subsection{Number-of-Objectives Based Classification}

\vspace{2mm}
\subsubsection{Single-Objective GAs}

Single objective GAs (SOGAs) are popular when it comes to optimize single objective with scalar fitness function. They are computationally less exhaustive but are required to run multiple times using different weight vectors to optimize multiple objectives as in case of multi-objective GAs \cite{ishibuchi2006comparison}. 


\vspace{2mm}
\subsubsection{Multi-objective GAs}

\begin{figure}[ht!]
\centering
\includegraphics[width=.28\textwidth]{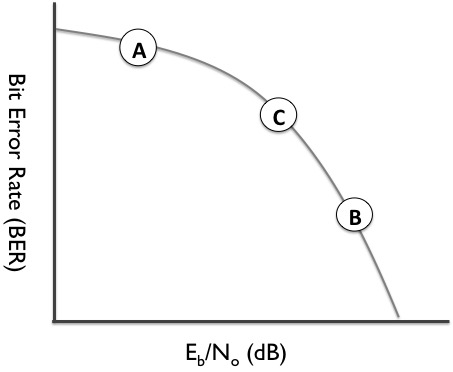}
\caption{Pareto front illustrated for two objectives}
\label{fig:Pareto}
\end{figure}

In a multi-objective optimization, there are multiple objectives that need to be  simultaneously optimized. In such a scenario, a single solution may not be suitable to multiple objectives. A solution may be best in one objective but worst in other cases. The overall system is optimized, in all directions, by multiple solutions. These categories of solutions are non-dominated solutions that lie on the \emph{Pareto front}. Each point in Pareto front is optimal, since no single solution exists that may improve one objective vector without degrading at least one other objective. Almost all the solutions lying on the Pareto front would be compromises relative to multiple objectives. There are various objectives in wireless networking that need to be optimized such as the signal to interference and noise ratio (SINR), the power consumption, and the bit error rate (BER). The Pareto front is illustrated through an example in Figure \ref{fig:Pareto} with two conflicting objectives: $A$ is a non-dominated point with the objective of minimizing power at the cost of BER, while the point $B$ corresponds to the objective of minimizing BER at the cost of power consumption. Both objectives cannot be achieved simultaneously so a compromise solution has to be adopted.

Multi-objective GAs (MOGAs) can be employed in almost every optimization setting   provided that chromosomes encoding and objective function are appropriately accounted for. Details and applications of MOGAs are discussed next in Section \ref{sec:applications}. In this section, we will describe one particular instance of a MOGA framework known as Non-dominated Sorting Genetic Algorithm (NSGA).

\vspace{2mm}
\paragraph{Non-dominated Sorted Genetic Algorithm (NSGA)}

NSGA is a class of multi-objective optimization algorithm that works at improving the adaptive fit of a population to the Pareto front constrained by a set of objective functions. NSGA can also be seen as a case of an evolutionary algorithm from the domain of evolutionary computation. The population within the algorithm is sorted into levels or sub-populations, which are based on the arrangement of the Pareto dominance. Major similarities between the population members of each sub division are calculated against the Pareto front, and the resulting class, along with its similarity measure, is then utilized to promote a solutions front that is non-dominated. Non-dominated Sorting Genetic Algorithm-II (NSGA-II) is a widely used computationally fast and elitist MOGA approach 
\cite{deb2002fast}. NSGA-II variants include the following:

\begin{itemize}

\vspace{1mm}
\item \emph{Steady State NSGA-II (ssNSGAII)}: A steady state version of the generational genetic algorithm. 

\item \emph{Parallel NSGAII (pNSGA-II)}: A variation of the NSGA-II wherein the algorithm takes advantage of the different cores of a computer in order to evaluate the individuals of a function in parallel. 

\item \emph{Random NSGAII (NSGA-IIr)}: NSGA-IIr is a variation of NSGA-II wherein the three basic operations including evolution, mutation and crossover are selected using a random seed to generate new individuals. 

\item\emph{Adaptive NSGA-II (NSGA-IIa)}: This approach works on the same principles as NSGA-IIr but the operators are adaptively selected through a random process.

\end{itemize}


\subsection{Other Types of GAs}

\vspace{2mm}
\subsubsection{Adaptive GA}

To prevent GAs from suboptimally converging to a local optima, a number of adaptive techniques have been recommended for adjusting parameters such as crossover probability, mutation probability, and population size. These techniques constitute adaptive GAs. In contrast to simple GAs that rely on fixed control parameters, adaptive GAs utilize information about the population in each generation to adaptively adjust the probability of crossover and mutation. 

\vspace{2mm}
\subsubsection{Parallel GA}

Parallel GA (PGA) can be categorized into three subcategories: \textit{(i)} master-slave GA; \textit{(ii)} island GA (also referred to as distributed GA or coarse-grained GA); and \textit{(iii)} cellular GA (also called fine-grained GA). Amongst these three configurations, the Island PGAs (island GA, or iGA, for short) are best suited to wireless networking since they allow control of the migration policy unlike the cellular PGAs---which require local communication at all agents during each generation of the PGA---and the master-slave PGAs that require the master and slaves to constantly communicate. 

\vspace{2mm}
\subsubsection{Distributed or Island GA}

Distributed GA divide a large population into several smaller subpopulations that interact weakly. Parallel GAs, on the other hand, are primed to speed up execution by implementing the sequential GA on several computation engines. Amongst the distributed GA approaches, the island GA approach is important and is popularly used in wireless networks \cite{friend2008architecture} \cite{elnainay2009channel}. It works by separating the population into sub-populations called islands. These islands cooperate with other islands through the migration of individuals. In island GA, the GA runs on each node with the possibility of good solutions migrating between nodes. These migrations are performed in accordance with the migration policy that controls the `when' and `how' of the individual movement. In the island GAs, there is relatively infrequent migration between island models thereby implying reduced inter-agent communication requirements compared to master-slave GA and cellular GA \cite{alba1999survey}.

\subsection{GA Components And Operators}

%
%
%

\begin{figure}[t]
\begin{center}
\includegraphics[width=.25\textwidth]{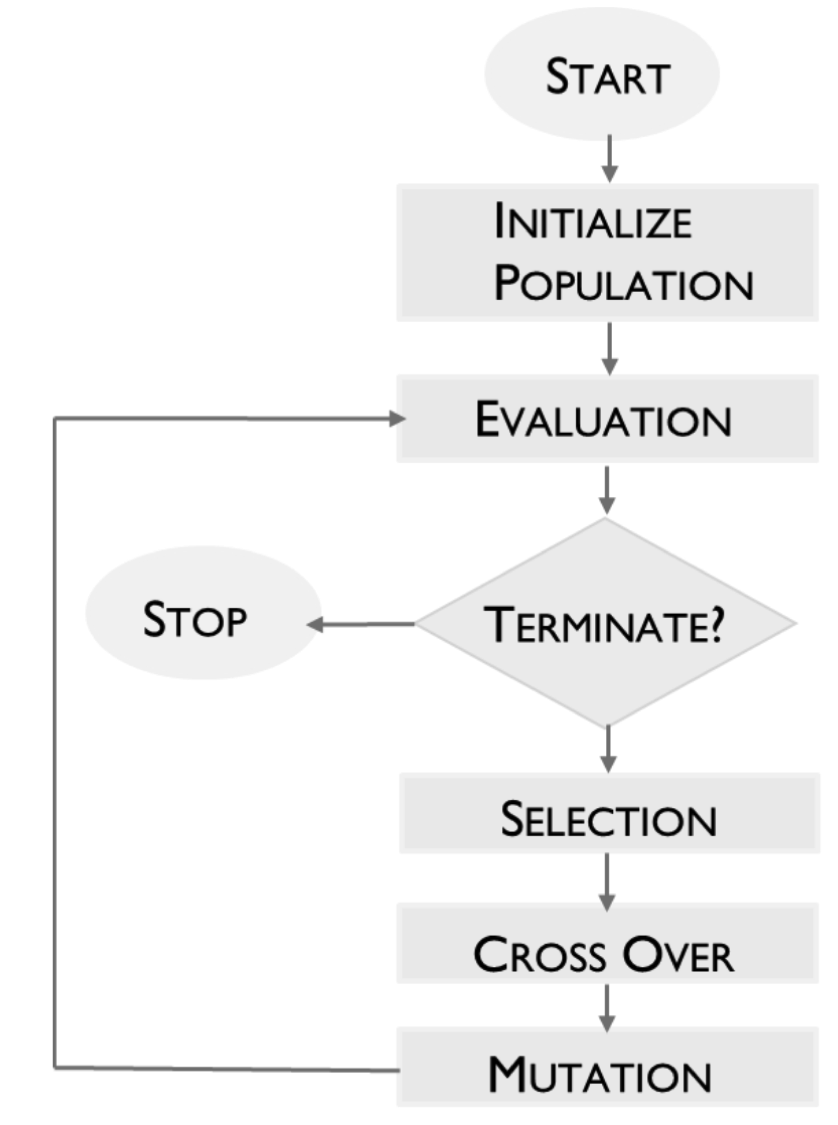}
\caption{A flow chart of a typical genetic algorithm \cite{qadir2013artificial}} 
\label{fig:ga}
\end{center}
\end{figure}

A flow chart of a typical GA is provided in Figure \ref{fig:ga}. The operation of a GA-based framework starts by initially defining a population of candidate solutions. While it is acceptable, and indeed commonplace, to define the population randomly, the convergence process can be accelerated by starting with an appropriately defined population. Each candidate solution defined in the population is known as an individual, with each individual encoded in an abstract format named as a chromosome (in a metaphorical reference to the encoding role of chromosomes in biological genetics). Various evolutionary techniques (mutation, crossover, etc.) are then applied on the population. In every generation, multiple individuals are stochastically selected from the current population with fitter individuals being more likely selections (due to the genetic principle of `survival of the fittest'). The selected individuals are then genetically modified (mutated or recombined) to form the next generation of the population. The usage of genetic operators and stochastic selection allow a gradual improvement in the `fitness' of the solution and allow GAs to keep away from local optima. 

In the following subsections, we will provide more details about the various steps involved in problem solving through GAs.

\vspace{2mm}
\subsubsection{Encoding into Chromosomes}

\begin{figure}[!ht]
\centering
\subfigure[\textbf{Binary Encoding.} Most common encoding type in which each chromosome is represented using a binary string.]{
 \includegraphics[width=.4\textwidth]{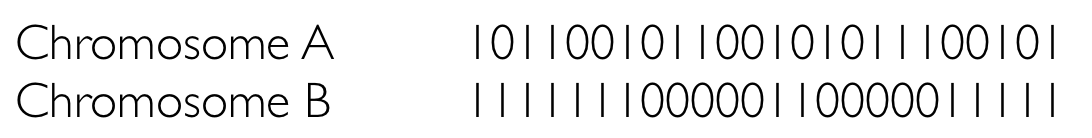}
 }
\subfigure[\textbf{Permutation Encoding}. Useful for ordering problems in which each chromosome represents position in a sequence.]{
   \includegraphics[width=.45\textwidth]{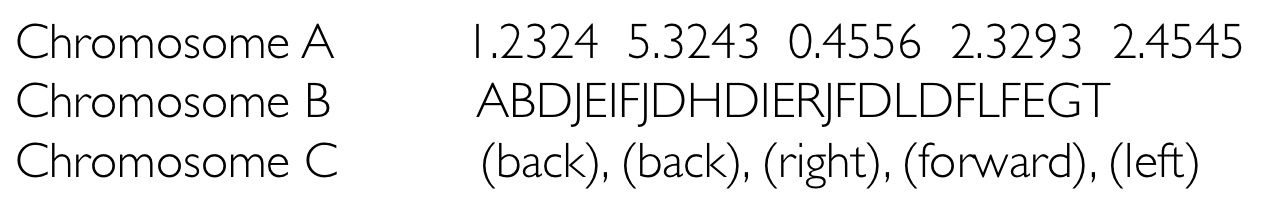}
}
\subfigure[\textbf{Value Encoding}. Chromosome are strings of some values (which can be form numbers, real numbers, chars, or some other complicated object).]{
   \includegraphics[width=.35\textwidth]{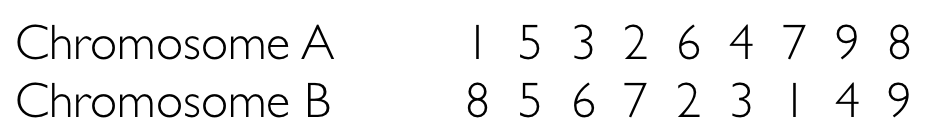}
}
\subfigure[\textbf{Tree Encoding}. Every chromosome is a tree of some objects (such as functions or commands in programming language).]{
   \includegraphics[width=.35\textwidth]{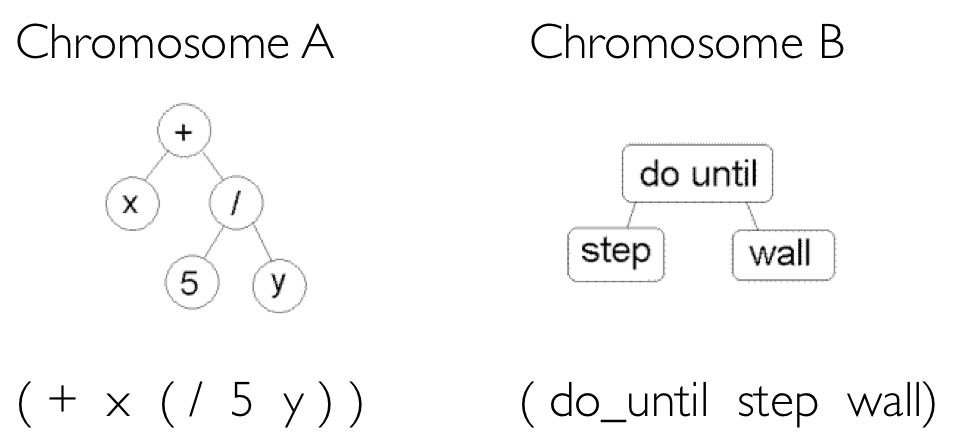}
}
\caption{Illustration of basic types of encoding schemes used by GA (Adapted from \cite{sivanandam442introduction}).}
\label{fig:encoding}
\end{figure}

Genetic algorithms operate by initially defining a \emph{population} of candidate solutions (each of which is called an \emph{individual}). Individuals are encoded in an abstract representation known as a \emph{chromosome} (which may be problem specific although representation in strings of 1s and 0s is common; in the related field of `\emph{genetic programming}', the encoding is in terms of snippets of computer code). The population size can be a significant factor in GA performance and efficiency. In particular, the performance of GAs can poor when the population size is very small \cite{grefenstette1986optimization}.


The encoding of candidate solutions in the form of chromosomes can also be of variable length. Such variable length encoding is well suited to problems such as shortest-path routing problem that compares paths of different lengths. As an example of variable length chromosome encoding, refer to the work of Ahn et al. \cite{ahn2002genetic} in which the routing path is encoded in the chromosome as a sequence of positive integers representing node IDs of the nodes enroute the end-to-end routing path. Each locus of the chromosome corresponds to the order of the node in the end-to-end routing path with the source node always represented by the gene of first locus. The length of the chromosome is variable but is upper bounded by $N$ (the total number of network nodes).

\vspace{2mm}
Several encoding schemes for GA exist including:

\begin{enumerate}

\item[a)] \emph{Binary Encoding}: Binary encoding is the most common approach, mainly because initial GA-based works used this encoding. In binary encoding every chromosome is a string of bits, i.e. 0 or 1. Binary encoding provides many possible chromosomes even for a small amount of alleles. On the other hand, this encoding is in some cases not natural for many problems and in other cases corrections need to be made after mutation or crossover.

\item[b)] \emph{Octal Encoding}: The chromosome is represented in octal numbers (0-7).

\item[c)] \emph{Hexadecimal Encoding}: The chromosomes are represented using hexadecimal numbers. (0-9, A-F).

\item[d)] \emph{Permutation Encoding}: Permutation encoding can be used in GA for ordering problems, such as task ordering problems, or the `traveling salesman' problem. In permutation encoding, each chromosome is a represented as a string of numbers, which is actually a number in a sequence. In some problems, for some types of mutation and crossover, corrections must be inculcated to leave the chromosome in a consistent form, viz. it should be corrected to ensure that it has a real sequence in it.

\item[e)] \emph{Value Encoding}: Direct value encoding is useful in problems that deal with complicated values (such as a real number). Use of binary encoding method in this case would be very difficult. In value encoding, every chromosome is a string of some values. Values can be in any arbitrary form including numbers, characters, or any other complicated object. While value encoding is a good approach for some problems, it often requires  
new crossover and mutation operators defined specifically for the problem. 

\item[f)]\emph{Tree Encoding}: Tree encoding is used for expressions or evolving programs, specifically for genetic programming (GP). In tree encoding, every chromosome or gene is in a tree form of some objects, like commands in programming language or a function. Programming language like LISP is mostly used for this, because genetic programs in it are encoded in this form and can be easily parsed as a tree structure, thereby facilitating mutation and crossover to be performed relatively easily.

\end{enumerate}

\vspace{2mm}
For illustrative purposes, an example binary encoding, permutation encoding, value encoding, and tree encoding is presented in Figure \ref{fig:encoding}.

\vspace{2mm}
\textit{Encoding example:} Consider an example encoding of radio parameters in chromosomes and genes for GAs. The inherent difficulty in such a representation is the large relative difference between radio parameters. For example, frequency range of 1 MHz to 6 GHz with step size of 1 Hz would require more than 30 bit genes, while modulation has only handful of techniques so it would require far less bits. Chromosomes can, therefore, utilize large number of bits for frequency while utilizing small number of bits for the modulation gene. To make GAs independent of which radio it is optimizing, the variable bit chromosomes representation is suggested by Scaperoth et al. \cite{scaperoth2006cognitive} in the context of  software defined radios (SDR).

\vspace{2mm}
\subsubsection{Fitness Function}

The fitness function, sometimes called the objective function, offers a mechanism for the evaluation of the individual chromosomes. Fitness function can be single objective, or multi-objective. The fitness function is devised in a way that the individual chromosome is its variable input parameter. Fitness function ensures that the chromosomes passed on to the next generation are not violating any constraints. It assigns a fitness score to each individual depending on the quality of the individual in terms of the optimization goals and objectives. In particular, for the fields of both genetic algorithms and genetic programming, each design solution is represented as a string of numbers referred to as the chromosome. In fitness evaluation, after each round of completed simulation or testing, $N$ worst design solutions are deleted, and $N$ new ones from the best design solutions are created. Each design solution thus needs to be awarded a figure of merit, which indicates how close it came to meet the overall specification. This is achieved by applying the fitness function to the simulation or test results obtained from that solution.

Genetic algorithms require much effort in designing a workable fitness function.  Though a computer might generate the final form of the algorithm, it takes a human expert to define the fitness function since an inefficient function can result in convergence to an inappropriate solution, or no convergence at all. Speed of execution of the fitness function is very important, since a typical GA needs to be iterated many times in order to produce a usable result for a non-trivial problem. Hence, fitness function requirements include:

\begin{itemize}
\item Minimum fitness computation time of candidate solution. 
\item Precise model definition for fitness computation.
\item Removal of noise from the fitness function values. 
\end{itemize}

Two main categories of fitness functions exist: one where the fitness function does not change over time (as in the case of optimizing a fixed function, or testing with a fixed set of test cases) and the other where the fitness function is mutable (as in the case of co-evolving the set of test cases or in niche differentiation). Another way of defining fitness functions is in terms of a fitness landscape that depicts the fitness level for each possible chromosome. Definition of the fitness function may not be straightforward in many cases. In some scenarios, it becomes very hard or impossible to come up even with a guess of what fitness function definition might be. This difficulty is solved by `\textit{interactive GA}', which outsources evaluation to external agents including humans. The function is calculated iteratively if the fittest solutions produced by GA are not in the desired range.  

There are various strategies for assignment of fitness values to the subject population. The \textit{`aggregation-based'} strategy sums the objectives into a single parameterized objective value, e.g., an aggregation of weighted-sum. The \textit{`criterion-based'} method transitions between the objectives during the selection phase of the algorithm. Every time an individual is chosen for reproduction, a potentially different objective will dictate which member of the population will be copied into the mating pool. In the `\textit{Pareto dominance based}' methods, different approaches like \textit{dominance depth}, \textit{dominance count}, and \textit{dominance rank} are used. The dominance rank dictates the number of individuals by which certain individual is dominated. In the dominance depth, the population is divided into several fronts and the dominance depth reflects the front to which an individual belongs. The dominance count is the number of individuals that are dominated by a certain individual.

\emph{Fitness function example}: The evaluation of the solution is at the heart of GAs since it is on the basis of fitness values that GAs evolve solutions and remove poor individuals. It can be thought of as the performance metric for the optimization algorithm. As an example, for the routing problem, the fitness can be equated with high average throughput ($f_i = t_i$: where $f_i$ is the fitness of i$^{th}$ individual and $t_i$ is its average throughput), low delay ($f_i = -d_i$) where $d_i$ is delay perceived by the i$^{th}$ individual. For classification problems, the fitness function can be some measure of the error. Realistic fitness functions are typically much more complex than this toy example, and can depend on many factors.

\vspace{2mm}
\subsubsection{Mutation}

\begin{figure}[ht!]
\centering
\includegraphics[width=.35\textwidth]{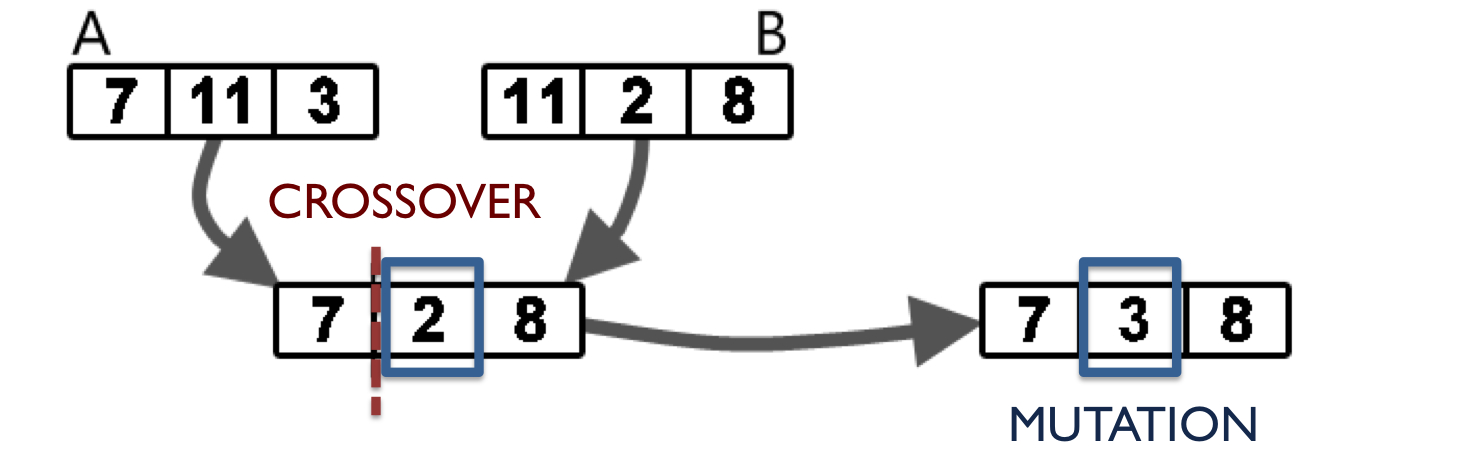}
\caption{Illustration of mutation and crossover operation. }
\label{fig:MutationCrossover}
\end{figure}

The mutation operator is used for \textit{exploration} with randomly altering genes in chromosomes to discover new horizons and new traits. Mutation helps to diversify the population, thus helping GAs to avoid local optima and pave the way towards global optima. Several types of mutation techniques exist including:

\begin{itemize}

\item \emph{Bit-string Mutation}: The mutation of bit strings occurs, through bit flips at random positions of chromosomes (as illustrated in Figure \ref{fig:MutationCrossover}).

\item \emph{Flip Bit}: This mutation operator takes the chosen genome chromosome and inverts the bits, i.e. if the genome bit is 0, it is changed to 1 and vice versa.

\item \emph{Boundary}: This mutation operator replaces the chromosome with either upper or lower bound in a random manner. This can be used for float and integer representation of chromosomes.

\item \emph{Non-Uniform}: The probability that the extent of mutation will go down to 0 with the next generation of chromosome is increased by using non-uniform mutation. This keeps the population from stagnating at early stages of the evolution process. It also alters solution in later stages of evolution. The mutation operator can only be used for float and integer chromosome genes.

\item \emph{Uniform}: This operator replaces the score of the chosen chromosome with a uniform random value selected in the range of the user-specified upper and lower bounds. This mutation operator can only be used for integer and float genes.

\item \emph{Gaussian}: This operator adds a unit Gaussian distributed random value to the selected chromosome. If it falls outside of the boundary of user-specified upper or lower bounds for that chromosome, the new value is clipped. This mutation operator is used for integer and float strings.

\end{itemize}

\vspace{2mm}
\subsubsection{Crossover}

The crossover process is essentially an attempt to \textit{exploit} the best traits of the current chromosomes and to mix them in a bid to improve their fitness. This operator randomly chooses a locus and exchanges the sub-sequences before and after that locus between two parent chromosomes to create a pair of offspring. A crossover point is randomly chosen. There can be multiple crossover points to aid more exploration. In biological systems, crossover is much more rampantly observed than mutation, often as much as a million times more frequent \cite{holland1995hidden}. Crossover along with mutation provides the necessary `evolutionary mix of small steps and occasional wild gambles' to facilitate robust search in complex solution spaces \cite{harford2011adapt}.

In GAs, crossover is viewed as synthesis of best practices and mutation is viewed as a method of spontaneous inspiration and creativity. The evolution can start from a population of completely random individuals and can evolve to better solutions through \emph{survival of the fittest} after application of genetic operators. The selected individuals (or chromosomes) are genetically modified (mutated or recombined) to form the next generation of the population.  The usage of genetic operators and stochastic selection allow a gradual improvement in the fitness of the solution and allow GAs to keep away from local optima. 



\begin{figure}[ht!]
\centering
\includegraphics[width=.45\textwidth]{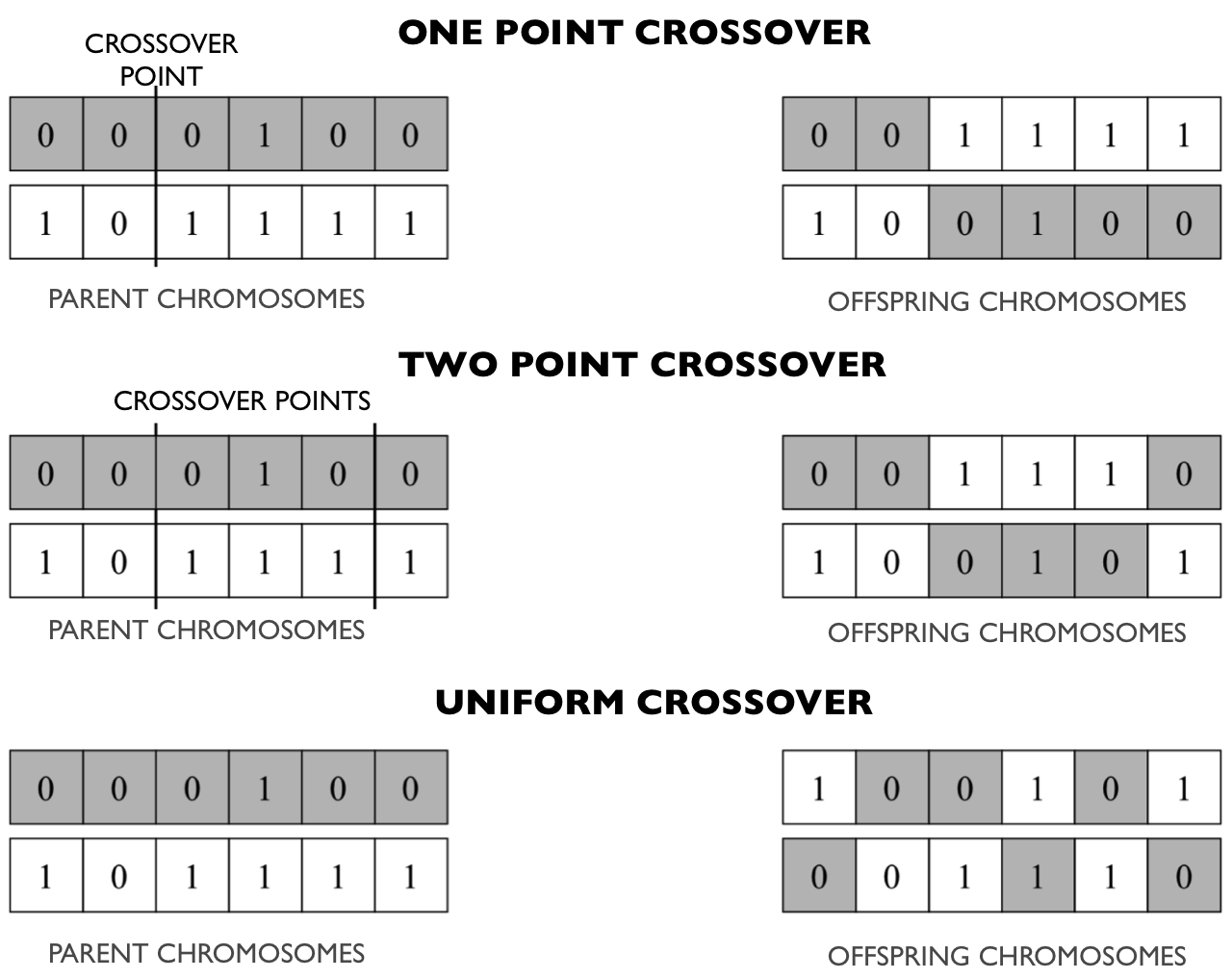}
\caption{Illustration of examples of one point, two points, and uniform crossover methods (Adapted from \cite{sastry2005genetic})}
\label{fig:Crossover2}
\end{figure}

Several methods exist for crossover of chromosomes which use different data structures \cite{goldberg1988genetic}. 

\begin{itemize}

\item \emph{One-Point Crossover}: A single crossover point on both parent's chromosome strings is selected. All data beyond that point, in either organism chromosome, is swapped between the two parent strings. The resulting strings are the new children chromosome. 

\item \emph{Two-Point Crossover}: Two-point crossover requires two points to be selected on the parent chromosome strings. Everything between the two points is swapped between the parent chromosome strings, producing two child chromosomes. 

\item \emph{Uniform Crossover and Half Uniform Crossover}: The uniform crossover approach uses a fixed mixing ratio between two parents strings. Unlike one-point and two-point crossover, this approach enables the parent chromosomes to contribute at the whole chromosome gene level rather than at the segment level. For example if the mixing ratio is 0.5, the offspring chromosome has approximately half of the genes from first parent and the other half from other parent, although cross over points can also be randomly chosen instead. In the half uniform crossover method, exactly half of the non-matching bits are swapped between the chromosomes. First the number of differing bits, or the \emph{Hamming distance}, is calculated. This number is then divided by two; the resulting number depicts the number of bits that do not match between the two parents, and will thus be swapped. An illustrative example comparing one point, two points, and uniform crossover is presented in Figure \ref{fig:Crossover2}. 

\item\emph{Cut and Splice}: In the crossover variant called the `cut and splice' approach,  a change in length of the children strings occurs. The reason for this difference in length is that each parent string has a separate choice of deciding crossover point.

\item \emph{Three Parent Crossover}: In this method, the child chromosome is derived from three parents which can be randomly chosen. Each bit of first parent is compared with bit of second parent to see if they are the same. If the bits are same then that corresponding bit is taken for the offspring otherwise the bit from the third parent is taken for the offspring chromosome. 

\item \emph{Ordered Chromosome Crossover:} In some cases, a direct swap may not be possible. One such case is when the chromosomes are in an ordered list. The N-point crossover can be used for ordered chromosomes, but always requires a corresponding repair process. Some of the approaches for ordered pair crossover include cycle crossover, position based crossover, alternating position crossover, edge recombination crossover, sequential constructive crossover and voting recombination crossover.

\end{itemize} 


\vspace{2mm}
\subsubsection{Selection}

Selection in GA allows individual genomes to be chosen from a population for later recombination or crossover (breeding). This operator performs the selection of chromosomes from population for reproduction. The fitter the chromosomes, the more likely they are to be passed on to the next generation. 
A number of selection techniques has been employed in GAs to carry out the survival of the fittest from a pool of individuals \cite{fette2009cognitive}.

A common selection procedure may be implemented in the following way:

\begin{itemize}

\item The fitness function is evaluated for each individual and is then normalized for algorithm use. The fitness value of each individual chromosome is typically normalized by the aggregation of all fitness values to ensure that  all the resulting fitness values sums up to 1.

\item The population is then sorted in descending order of fitness values.

\item Accumulated normalized fitness scores are calculated. Accumulated fitness value of an individual equals the sum of fitness scores of all the previous individuals and its own fitness score. Accumulated fitness of the previous individual must always be 1---Any other value can be used as indication of a wrong normalization procedure.

\item A random number $R$ is chosen in the range 0 to 1. The first individual whose summed up normalized value is greater than the number $R$ is selected.

\end{itemize}

This selection method is called fitness proportionate selection or roulette-wheel selection, since the procedure is repeated until there are enough selected individuals. If instead of using a single pointer that is spun a number of times, multiple equally spaced pointers are used on a wheel that is spun once, the selection process will be called `\textit{stochastic universal sampling}'. The selection process will be a tournament selection process once there is repeated selection of the best-fitted individual from a randomly chosen subset pool. This will be further called truncation selection once we take the best half, third or some other proportion of the individuals.

There are also other selection mechanisms that do not take into account all individuals for the selection process, but only those that have a fitness value, greater than a given constant. Other algorithms select the individual from a restricted pool where only a constrained percentage of the individuals are allowed, based on their fitness scores. Retaining the best individual chromosomes in a generation unaltered, in the next set of generation, is called elitist selection or \emph{elitism}. It is considered a successful variant of the general procedure of constructing a new population pool.

The main selection techniques used by GAs are summarized in the following list.

\begin{itemize}

\vspace{1mm}
\item \emph{Relative Tournament Evaluation}: In relative tournament, two parents are compared randomly and the one with high fitness value is selected to be the parent of next generation. Winning chromosomes are awarded by adjusting their weight corresponding to the objective function. This weighting increases the importance of that chromosomes relative to others.

\vspace{1mm}
\item \emph{Roulette Wheel Selection}:
Roulette wheel selection is a way of choosing mother and father chromosomes from the population in a way that is proportional to their fitness. The chromosomes with higher fitness have high chances of being selected using this technique.

\vspace{1mm}
\item \emph{Relative Pooling Tournament Evaluation}: In this technique, each parent is thrown in a competition with the individual chromosomes in population, independently. The individual that is the winner of maximum number of tournaments gets a chance to be passed into next generation through  reproduction. This technique is comparable to the proposed method in Horn et al.\cite{horn1994niched} of fighting two individual against various subsets of population. This technique is computationally intensive, and thereby can stress GA deployment due to the computational overhead.

\vspace{1mm}
\item \emph{Elitism}: Elitism is a method to store and use previously found best-suited solutions in subsequent population generations of GA. In a GA approach that uses elitism, the statistic related to the population's best solutions does not degrade with generation. Maintenance of archives for non-dominated solutions is an important and critical issue. The final contents of archive usually represent the result returned by applying the optimization process. It is highly effective and common to employ the archive as a pool, from which guidance is taken for the generation of new solutions. Some algorithms use solutions available in the archive exclusively for such purpose, while others tend to rely on the archive to varying degrees according to settings of related parameters. The computational complexity of maintaining the archive which involves checking for non-dominance of newly generated solutions suggests relatively modest size of bounded archive \cite{knowles2004bounded}. 

\end{itemize}

\section{Applications of GAs in Wireless Networking}
\label{sec:applications}

After providing background on GAs in the last section, we are now in a position to survey the applications of GAs in wireless networking. GAs have been applied in wireless networking in a wide variety of contexts. We will firstly describe the various applications of GA in wireless networking in Section \ref{sec:appbased}. We will then provide a network-type based classification of GA application in Section \ref{sec:networkbased}. Finally, we will discuss `hybridization' proposals that combine GAs with other techniques in Section \ref{sec:Hybrid}.

\subsection{\textbf{Application Based Classification}}
\label{sec:appbased}

\begin{table}
\centering
\caption{Applications of genetic algorithms in wireless networking (covered in Section \ref{sec:appbased})}
\begin{tabular}{p{3cm}p{4cm}}
\toprule
 \textbf{\emph{Application}} &  
 \textbf{\emph{Discussed in}} \\
\midrule

\multicolumn{1}{l}{\textbf{\emph{Routing}}} &  
Section \ref{sec:routing} and Table \ref{tab:Routing}\\

\multicolumn{1}{l}{\textbf{\emph{Quality of Service (QoS)}}}  &  
Section \ref{sec:QoS} and Table \ref{tab:QoS}\\
 
 \multicolumn{1}{l}{\textbf{\emph{Load balancing}}}  &  
 Section \ref{sec:LoadBalancing} and Table \ref{tab:LoadBalancing}\\

 \multicolumn{1}{l}{\textbf{\emph{Bandwidth allocation}}} &
 Section \ref{sec:BandwidthAllocation} and Table \ref{tab:bandwidth}.\\

 \multicolumn{1}{l}{\textbf{\emph{Channel allocation \& assignment}}} &
 Section \ref{sec:channelAA} and Table \ref{tab:channelalloc}.\\
  
  \multicolumn{1}{l}{\textbf{\emph{Localization}}} & 
  Section \ref{sec:Localization} and Table \ref{tab:Localization}\\
  
    \multicolumn{1}{l}{\textbf{\emph{Wireless AP placement}}} &  
    Section \ref{sec:APplacement}  and Table \ref{tab:AP_placement}
\\

    \multicolumn{1}{l}{\textbf{\emph{General Applications}}} &
    Section \ref{sec:GeneralApplications} and Table \ref{tab:generalApplications}
    \\

\bottomrule 
\end{tabular}
\label{tab:summary}

\end{table}

In this section, we will discuss the applications of GAs in wireless networking. In particular, we will discuss routing, QoS, load balancing, channel allocation and assignment, localization, WLAN AP placement, and other general applications, in this particular order. The organization of this section is summarized in the Table \ref{tab:summary}.

\vspace{2mm}
\subsubsection{Routing}
\label{sec:routing}



\begin{table*}
\caption{Using genetic algorithms for \textit{\underline{routing}} in wireless networks}
\centering
\begin{tabular}{p{2cm}p{0.8cm}p{1cm}p{3.7cm} p{8.2cm}}
\toprule
 \textbf{\emph{Project \& \newline reference}}   &  \textbf{\emph{Network}}  &  \textbf{\emph{No. of \newline objectives}}  &  \textbf{\emph{Technique used}} & \textbf{\emph{Brief summary}} \\
\midrule

Ahn and Ramakrishna \cite{ahn2002genetic}  & MANET & SOGA & Integer coded, variable length chromosomes &  Unlike Dijkstra and Depth-First Search (DFS) algorithms, multiple solutions can be found for shortest path routing that have same cost using GAs.\\

Cheng et al. \cite{cheng2010multi}  & MANET & MOGA & Distributed multipopulation GA &  A distributed GA approach incorporating multi-population with random immigrant schemes solves the shortest path routing in MANET.\\

Gen et al. \cite{gen1997genetic}  & LANs & SOGA & Integer coded GA for random distributions &  A priority encoding scheme is proposed using GAs to represent and optimize all possible paths in network graph.  \\

Davies et al. \cite{davies2003genetic}  & WAN & SOGA & Simple GA with unbiased coding scheme &  A GA approach continuously re-routes and assigns weight to traffic paths to implement adaptive control over dynamic routes and network topology.\\

Yang et al. \cite{yang2010genetic}  &  MANET  & SOGA &  GA with immigrants and memory scheme &  A GA proposed for investigating effectiveness and efficiency of GAs with immigrants and memory schemes in solving the dynamic shortest path routing problem in the real-world networks.\\

Badia et al. \cite{badia2009genetic}  &  WMNs  & SOGA & ILP formulation with GAs &  GA addressed the routing and link scheduling problem when ILP cannot cope with computational complexity of large WMNs.\\

Al-Karaki et al. \cite{al2009data}  &  WSN  & MOGA & Centralized GA and ILP &  A metaheuristic algorithm that solves the joint problem of optimal routing with data aggregation to optimize the network lifetime while retaining the energy and latency efficiency. \\

Lorenzo and Galisic. \cite{lorenzo2013optimal}  &  CN  & MOGA & Sequential GA &  A sequential GA solves the NP-hard problem by optimizing relay topology while eschewing the inter-cell interference.\\

Apetroaei et al. \cite{apetroaei2011genetic}  & WSN & SOGA  & Tree coded GA on spanning tree topology &  GA inspired spanning tree topology for WSNs that adapts (according to the nodes, residual energy) to maximize the usage of the network.\\

Yen et al. \cite{yen2008genetic}  & MANET  & SOGA & Pr\"ufer number encoding &  A GA strategy for energy efficient routing by maintaining the energy balanced nodes in entire MANET distribution.\\

Chiang et al. \cite{chiang2007near}  &  MANET  & SOGA & Tree sequenced coding &  A GA based on sequence and topology encoding employed topology crossovers to solve the multicast routing constraints efficiently.\\

\\ \bottomrule
\end{tabular}
\label{tab:Routing}
\end{table*}

The problem of \textit{shortest path routing problem} is a well-studied problem in literature. GAs are well suited to the routing problem allowing convenient mechanisms for developing adaptive routing solutions. GAs have been used for developing routing solutions in a variety of wireless settings as summarized in Table \ref{tab:Routing}. Ahn et al. \cite{ahn2002genetic} have tackled this problem  through GAs. The routing table for each source-destination pair needs to be constructed first. Obviously, the number of possible routes between two nodes heavily depends on the network topology. If the network is densely connected or the size of the network is large, the number of possible routes of a source-destination pair becomes huge. Hence, it is impossible to list all the possible paths in the routing table. A predefined number of routes are randomly generated and fed to population pool, and chromosomes are then optimized. In the simulation, crossover and mutation are run on \emph{variable length} chromosomes. Infeasible solutions are also kept in check. The paper discussed the issues of  path-oriented encoding, and path-based crossover and mutation, which are relevant to the routing problem.


Yang and Wang \cite{yang2010genetic} studied how GAs can be used for \textit{dynamic shortest path routing problem} (DSPRP). Algorithms like Bellman-Ford and Dijkstra perform well in fixed infrastructure but on dynamic scale, high computational complexity renders them unacceptable. A DSPRP model is built up in this paper using GAs. A specialized GA is designed for the shortest path problem in MANETs. Several immigrants and memory schemes that have been developed for GAs for general dynamic optimization problems are adapted and integrated into the specialized GA to solve the DSPRP in MANETs. The experimental results indicate that both immigrants and memory schemes enhance the performance of GAs for the DSPRP in MANETs.

In MANETs, all the nodes communicate over wireless links without any established infrastructure. MANETs are well suited to environments in which there is no fixed infrastructure or when the infrastructure is not trusted. In such networks, a common strategy is to form clusters where each node is linked to a clusterhead for efficient routing with other nodes that are not in its direct range. GAs have been used in such cluster-based routing schemes for MANETs. Al Gazal et al. \cite{al2007routing} have proposed a GA-based protocol named `\textit{cluster gateway switch routing protocol}' (CGSRP) to select the clusterhead to carry out communication between nodes. Clusterhead must have enough resources, power, and bandwidth to avoid dangers of bottlenecks. This scheme works by encoding each node's unique ID in the chromosomes. The chromosomes have information about cluster-head, members, number of links in each cluster-head. The encoded chromosomes are then evaluated against certain criteria as defined by the fitness function (which may incorporate features such as load balancing and bandwidth conservation). Each chromosome's fitness is then evaluated. The process of the survival of the fittest leads to an optimum selection of nodes as cluster-heads that optimally utilize resources. 


GAs have also been proposed for use in multicast and broadcast routing in wireless networks. A study on multicast routes optimization using GAs is done by Banerjee and Das \cite{banerjee2001fast}. An on demand routing scheme with GAs is discussed to meet bandwidth, delay constraints while maintaining the QoS in wireless networks. Chromosomes from different pools are randomly concatenated to obtain a multicast solution. In doing so, all possible routes between source and destination nodes are encoded and fed to GA in the form of population chromosomes. GA then uses evolutionary techniques to select the best chromosomes by evaluating them against a particular fitness function, which can be in the form of different constraints related to bandwidth, QoS, and channel allocation, etc. In another work, Salcedo-Sanz et al. \cite{salcedo2003mixed} proposed mixed neural-genetic algorithm for the `\textit{broadcast scheduling problem}' (BSP). The algorithm solves this optimization problem in two stages: firstly, it finds a feasible solution, and secondly it obtains maximal transmission throughput by employing fittest chromosomes through GAs. A track of interference and frame matrix are kept and the channel slots are assigned accordingly. Nodes that are 1-hop away are not assigned the same slot to avoid mutual interference. After finding feasible solutions, these solutions are encoded in chromosomes. Multiple GA operators---such as mutation, crossovers---are then applied leading to maximized throughput.

The use of hybrid GA algorithms have been proposed for routing in wireless networks. Cauvery et al. \cite{nkrouting} proposed using ant algorithms, combined with GAs, for route discovery. It was observed that the use of ant algorithm leads to reduction in the size of the routing table. We note here that ant-colony-optimization (ACO) algorithms belong to the broader class of AI known as swarm intelligence. While GAs cannot use global information of the network, the use of ACO and GA techniques in tandem leads to independent exploration of the network helping in effective computation of routes between any pair of nodes. The proposed algorithm creates an initial population, determines the forward-ants, generates new populations using genetic operators and fitness evaluation, and finally updates routing table. The hybrid algorithm performs well in not only generating routes between pair of nodes but also for achieving efficient load balancing.

\vspace{2mm}
\subsubsection{Quality of Service (QoS)}
\label{sec:QoS}

\begin{table*}
\caption{Using genetic algorithms for \textit{\underline{QoS optimization}} in wireless networks}
\centering
\begin{tabular}{p{1.8cm} p{3cm} p{1.1cm}p{2.8cm} p{7.5cm}}
\toprule
\textbf{\emph{Reference}}  &  \textbf{\emph{Goal}}  &  \textbf{\emph{No. of\newline objectives}}  &  \textbf{\emph{Technique used}}  &  \textbf{\emph{Brief summary}} \\
\midrule

\multicolumn{3}{l}{\textbf{\emph{Resources Scheduling for QoS }}} \\
Sherif et al. \cite{sherif2000adaptive}  & Resources allocation with call admission control & SOGA & GA with reallocation of unused bandwidth &  Accelerate allocation through crossover and mutation. Reduces computation of NP-hard problem. Dynamic assignment of network resources and bandwidth\\

Gozupek and Alagoz \cite{gozupek2011genetic}  &  Scheduling under interference constraints & MOGA & Binary encoding with random seeding  &  GA based sub-optimal schedulers for maximizing the throughput and improving the latency in CRNs.\\

Cheng and Zhuang \cite{cheng2009novel}  & Packet level resource allocation  & SOGA & Karush-Kuhn Tucker (KKT) and GA driven approach &  An effectual hybrid GA optimizes the QoS provisioning and resource allocation to keep a balance between performance and computation.\\

Lopez et al. \cite{lopez2014genetic}  &  Transmit power control for maximum utility  & MOGA & GA with call admission control &  MOGA approach adopted to maintain the transmit power of CRs, while maintaining the QoS and optimizing the spectrum access.\\

Roy et al. \cite{roy2002efficient}  & Energy efficient resource scheduling & MOGA & Depth first search with NSGA &  On demand QoS routing scheme has been proposed based on end user needs. A solution meeting the constraints is offered from an optimized pool of chromosomes.\\

\multicolumn{3}{l}{\textbf{\emph{QoS for Multimedia and Web-Services }}} \\
Jiang et al. \cite{jiang2011multi}  &  Multi-Constrained QoS for web-services   & SOGA & Variable length chromosome GA & A scalable QoS aware web service composition is introduced using GA for diversified functionality through cut and splice operation of mutation and crossover.\\

Canfora et al. \cite{canfora2005approach}  & QoS service composition & SOGA & GA with ILP and elitism &  GA approach for web based QoS services with non-linear aggregate functions while meeting a set of constraints.\\

Li et al. \cite{li2006multi}  &  Video On Demand  & MOGA & GA with interference model & GA is employed to leverage multi-source, multipath solution to design a concurrent video streaming environment in WMNs, along with route discovery.  \\

\multicolumn{3}{l}{\textbf{\emph{Energy-Efficient QoS Multicast Routing}}} \\

EkabtaniFard et al. \cite{ekbatanifard2010multi}  &  Energy Efficient QoS routing in clusters   & MOGA & NSGA-II &  A hierarchical energy efficient routing protocol, for two tier WSN networks, under tight QoS constraints, is developed using MOGA optimization with NSGA-II. \cite{deb2000fast}.\\

GAMAN  & Topological Multicasting  & SOGA & Tree coded GA with multi-population &  A GA for MANETs (GAMAN) is proposed by Barolli et al. \cite{barolli2003qos}. The algorithm works locally on smaller population and offers promising QoS for ad-hoc networks. \\

QM$^2$RP \cite{roy2004qm2rp} & QoS-based multicast routing & MOGA & Tournament selection and niched Pareto optimization & A QoS-based mobile multicast routing protocol using multi-objective genetic algorithm.\\

Cui et al. \cite{cui2003multiobjective}  &  Optimizing conflicting QoS parameters & MOGA & Depth-first search algorithm and GA &  A MOGA approach is used to optimize multiple multicast routes lying on Pareto front.\\

Lu et al. \cite{lu2013genetic}  &  Energy efficient QoS multicast routing  & MOGA & Tree structured coding &  Proposed GA solves QoS by constructing a multicast tree with a minimum energy cost and bounded end-to-end delay.\\

Yen et al. \cite{yen2011flooding}  &  Multicast routing through flooding & MOGA & Tree coded flooded GA  &  A GA based method is proposed for multi-constrained QoS multicast routing, while incorporating elitism, to optimize the computation time.\\

Huang and Liu \cite{huang2010moeaq}  &  QoS Aware Multicast Routing  & MOGA & Greedy and family competition with GA &  A hybrid GA along with greedy algorithm mitigate premature convergence problems to avoid local optimas.\\

Sun et al. \cite{sun2008multiple}  &  Multi-constrained optimization & MOGA & Spanning tree encoding &  The evaluated GA can optimize the cost of the multicast tree, maximum link utilization, the average delay selection of the long-life path.\\

Abdullah \cite{abdullah2010multiobjectives}  &  Modified QoS Routing Protocol  & MOGA & Variable length encoding &  A GA approach  look for the best QoS route, using route mutation, to optimize the design of MANET routing protocol by improving packet delay and throughput.\\
\\
\multicolumn{3}{l}{\textbf{\emph{Route Discovery}}} \\
Gelenbe et al. \cite{gelenbe2006genetic}  & Autonomous route discovery with GA & SOGA & GA with distributed CPN protocol & 

Proposed autonomous route discovery in cognitive packet networks (CPNs) with GAs. Paths discovered by CPN are provided to GA which then produces new routes using evolutionary techniques\\

Cauvery and Visvanatha \cite{nkrouting}  & Route discovery through mobile agents  & SOGA & GA with Ant algorithm &  A hybrid algorithm (combining ant algorithms and GA) is proposed to reduce the size of routing table and help in routing among autonomous peers. \\
\\
\multicolumn{3}{l}{\textbf{\emph{Queue Management at Routers (AQM)}}} \\
Fatta et al. \cite{di2003genetic}  & Designing FPI controller for AQM & SOGA & GA with fuzzy logic &  Fuzzy proportional integer (FPI) controller is purposed and FPI parameters are controlled by GAs dynamically to optimize active queue management (AQM) policies. \\

Selamat et al. \cite{selamat2004routing} & Query Retrieval for active AQM & SOGA & Integer coded distributed GA &  A GA  with mobile agents search system (MaSS) is proposed to minimize the query retrieval cost while maintaining a reasonable path delay.\\

\bottomrule
\end{tabular}
\label{tab:QoS}
\end{table*}

\begin{table*}
\caption{Using genetic algorithms for \textit{\underline{load balancing}} in wireless networks}
\centering
\begin{tabular}{p{1.8cm} p{1.5cm} p{4cm} p{9.7cm}}
\toprule
\textbf{\emph{Reference}}  &  \textbf{\emph{No. of \newline objectives}}  &  \textbf{\emph{Technique used}} & \textbf{\emph{Brief summary}} \\
\midrule

\multicolumn{3}{l}{\textbf{\emph{Wireless Mesh Networks}}} \\

Hsu et al. \cite{hsu2008survivable}  & SOGA & GA with Dijkstra algorithm &  A hybrid GA coupled with Dijkstra's algorithm look for the best available paths with bounded delays and low cost network configuration. \\

Zeng et al. \cite{zeng2008load}  &  SOGA & GA with greedy algorithm and binary encoding &  Load balanced disjoint clustering is done by a GA coupled with \textit{greedy algorithm} for efficient distribution of network resources. \\

\\
\multicolumn{3}{l}{\textbf{\emph{Mobile Ad-hoc Networks}}} \\

Yen et al. \cite{yen2008genetic}  &  MOGA  & GA with Pr\"ufer number coding &  In time-critical load balancing scenario, GAs inspired strategy is applied in multicast tree to balance the energy consumption, leading to prolonged network lifetime.\\

Cheng et al. \cite{cheng2013dynamic}  &  SOGA & Distributed GA with multi-population and immigrant schemes &  A dynamic GA evaluates the fitness of each feasible clustering structure against load balancing metric.\\

\\
\multicolumn{3}{l}{\textbf{\emph{Wireless Sensor Networks}}} \\
He et al. \cite{he2012load}  &  SOGA & 1-to-1 mapping of nodes to genes in chromosomes &  The application of GA to find a load-balanced connected dominating set (CDS) to be used as a virtual backbone of WSNs.\\

\\
\multicolumn{3}{l}{\textbf{\emph{Wireless Local Area Networks}}} \\
Scully and Brown \cite{scully2009wireless}   & SOGA & MicroGA \&  MacroGA &  A comparison of MicroGA and MacroGA is done against QoS attributes like throughput and load balancing. MicroGA performs better in brisk load balancing under time constraints. \\

\bottomrule
\end{tabular}
\label{tab:LoadBalancing}
\end{table*}

With increasing prevalence of multimedia applications (such as video conferencing and streaming), and the development of high-speed wireless communication technologies, efficient and reliable QoS provisioning has become increasingly important. Modern multimedia applications place rigorous QoS demands for different parameters like bandwidth, delay, and jitter \cite{malik2014qos}. This motivates the development of effective resource allocation schemes that can ensure QoS while also satisfying these strict QoS constraints. In this section, we will describe the various GA-based QoS solutions that have been proposed in literature. This information is presented in summarized form in Table \ref{tab:QoS}.

Various analytic approaches have been proposed in literature including those that rely on Semi-Markov decision process (SMDP) theory for optimal call admission control (CAC) \cite{choi2000call}. These methods unfortunately fail to scale to large-scale networks where the decision and action space is large. GAs can provide near optimal solutions in large-scale complex networks. 
Xiao et al. \cite{xiao2000near} presented an effective near optimal GA-based CAC approach for large-scale wireless/ mobile networks. The chromosomes encoded are binary strings where 1 and 0 represent the acceptance and rejection of calls by CAC. These chromosomes are randomly generated and then evaluated against fitness function of channel utility and constraints. Evolution of best traits leads to near optimal solutions for QoS. This scheme offers adaptive QoS to multi-class users while being computationally less intensive compared to the SMDP approach.

A multi-objective GA (MOGA) approach on multicast QoS routing is proposed by Roy et al. \cite{roy2002efficient}. To deliver on-demand QoS, different solutions are kept in an optimized pool. Multiple solutions are offered depending upon the service demands. A MOGA approach optimizes the prioritized objectives as stated in the user QoS requirements. A similar approach for QoS aware web-based service under strict constraints and optimizing the targeted QoS parameters is done by Canfora et al. \cite{canfora2005approach}. It is shown in this study that the widely adopted optimization approach of \emph{linear integer programming} is unable to offer non-linear QoS functions. GAs can be employed to work with non-linear QoS attributes. In yet another MOGA work, Roy et al. have proposed QM$^2$RP as a QoS-based multicast routing protocol framework for mobile wireless cellular networks \cite{roy2004qm2rp}. The proposed protocol determines near-optimal routes on demand while working with potentially inaccurate network state information. The protocol considers a number of QoS optimization metrics including end-to-end delay, bandwidth, and residual capacity. 

With the increase in number of service candidates, the traditional chromosome model for QoS optimization becomes cumbersome in web services. To address this, Gao et al. \cite{gao2007qos} studied a `tree based GA' for optimizing QoS. This model is faster than traditional GA since it offers impulsive chromosomes coding and decoding and present run time integration of QoS resources. Chromosomes are represented in series of nodes and all the leaf nodes execute the process. A tree characterizes the structured model for different QoS composition, and requires less computation for large networks. Sub-tree crossover exchanges the task nodes and updates the QoS vector from the leaf nodes to the tree's root. A standardized and comprehensive fitness function verifies the whole QoS process plan. Results have showed that the tree based models have lesser overhead than the traditional GAs with serial chromosomes while offering excellent replanning and service composition.

\begin{table*}
\caption{Using genetic algorithms for \textit{\underline{bandwidth allocation}} in wireless networks}
\centering

\begin{tabular}{p{2.5cm} p{1cm}p{3cm}p{10cm}}
\toprule
\textbf{\emph{Reference}}    &  \textbf{\emph{No. of \newline objectives}} &  \textbf{\emph{Techniques used}}  &  \textbf{\emph{Brief summary}} \\
\midrule

Lin et al. \cite{lin2012accelerated} & MOGA & GA with adaptive inheritance probability & A prioritized bandwidth allocation scheme for transmitting medical patients data is applied for multimedia sub-streams.\\

Kandavanam et al. \cite{kandavanam2010hybrid} & MOGA & GA with variable neighborhood search approach & A hybrid GA optimizes the residual bandwidth allocation to routing links for multiclass traffic in all networks.\\

Karabudak et al. \cite{karabudak2004call} &SOGA& GA with Markov Decision Model and Elitism& The proposed GA scheme minimizes handoff latency by using elitism technique to allocate the maximum network resources to admitted calls.\\

Vendanthan et al. \cite{vedantham1998bandwidth}& SOGA & GA with greedy algorithms & A GA approach to efficiently utilize the bandwidth for different incoming call streams to maximize the revenue generation.\\

Riedl \cite{riedl2002hybrid} &SOGA& GA with local search heuristics & GA approach with bandwidth and delay metrics optimizes the IP traffic engineering in wireless networks.\\

Kobayashi et al. \cite{kobayashi2004designing}& MOGA &Distributed GA & A distributed GA works by utilizing both local and global optimization in parallel on networks nodes to utilize the network resources and bandwidth.\\

Qahtani et al. \cite{al1998dynamic} &SOGA&GA for asynchronous transfer mode (ATM)&A genetic routing algorithm (GRA) is designed to allocate bandwidth to virtual path subnetworks in a wireless network.\\

\bottomrule
\end{tabular}
\label{tab:bandwidth}
\end{table*}

\vspace{2mm}
\subsubsection{Load Balancing}
\label{sec:LoadBalancing}

GAs have been proposed for use in various load balancing solutions for various wireless networking configurations in literature (as shown in Table \ref{tab:LoadBalancing}).

There are three main types of load balancing categories: \emph{(i)} strongest `received signal strength indicator' (RSSI); \emph{(ii)} least loaded first (LLF); \emph{(iii)} hybrid load balancing (HLB) incorporating RSSI \&  LLB. These techniques fall under the category of network-centric load balancing, and are applied on a network scale to achieve global optima, unlike user-centered techniques that are more likely to achieve local optima only. Chromosomes consist of randomly assigned individuals to an access point. Crossover and mutation are performed on individuals and fitness evaluation is done. Both load balancing and congestion control problems have been addressed. QoS has also been addressed by assigning different weight to users, depending on its service requirement like streaming and media etc. Experimental results and analysis showed that microGA outperforms macroGA at first but in the long run macroGA does better load balancing by pushing more congested users towards peripheral APs.

Ozugar et al. \cite{ozugur2001multiobjective} proposed the multi-objective hierarchical mobility management optimization for UMTS networks to distribute the load among different parts of network such as serving GPRS nodes (SGSN), radio network controllers (RNC), and mobile switching centers (MSC) in addition to performing the role of minimizing cost for location update in these intricate networks.

Selecting resource-rich nodes for trivial tasks creates an imbalance in energy distribution among nodes resulting in the demise of multicast service. To mitigate such a possibility, Yen et al. \cite{yen2008genetic} have proposed an energy-efficient GA routing in MANETs. The algorithms proposed in \cite{yen2008genetic} continuously observe the condition at various nodes and improve the energy balance by replacing the weaker nodes with energy-rich nodes. An extended \emph{Pr\"ufer encoding scheme} is proposed for energy efficient exchange of leaf nodes through mutation and crossover. In this way a substantial reduction in convergence time is achieved, thus prolonging the service duration efficiently.

GAs have also been proposed for load balancing at APs in IEEE 802.11 WLANs for efficient distribution of bandwidth among users. Scully and Brown \cite{scully2009wireless} have discussed MicroGAs and MacroGAs to address the problem of load balancing. A comparison of MicroGAs and MacroGAs has been done by Kumar et al. \cite{krishnakumar1990micro}. MicroGAs has a population of five individuals leading to brisk evolution.

\vspace{2mm}
\subsubsection{Bandwidth Allocation}
\label{sec:BandwidthAllocation}

The problem of bandwidth allocation under QoS constraints in wireless networks is typically a complex task. A proper bandwidth allocation scheme must be adopted to overcome the problem of limited wireless resources and utilize the channel allocation efficiently. Various GA-based approached have been proposed in literature as solutions to the problem of bandwidth allocation in wireless networks. We will describe a sample of these works (a tabulated summary is presented in Table \ref{tab:bandwidth}.

An accelerated GA approach with dynamic adjusting of mutation and inheritance probability is proposed in \cite{lin2012accelerated} as a solution to the bandwidth allocation problem in WLANs. The probability of mutation and crossover is made dependent on the individual chromosomes fitness. This approach is shown to quickly converge to global optima while substantially reducing the number of generations as well as the computation time. In another work, Sherif et al. \cite{sherif2000adaptive} proposed a GA-based approach to analyze the network bandwidth allocation using a predetermined range of QoS levels (high, medium, low) declared by each multimedia substream. The proposed algorithm assigns the best available QoS level to each substream depending upon the availability of bandwidth resources in network. In the case of network congestion, the algorithm tries to preserve resources by degrading the quality of some admitted calls. Using this technique, the algorithm attempts maximum resource utilization and fair distribution of bandwidth in all the networks while keeping in view the network constraints.

\vspace{2mm}
\subsubsection{Channel Allocation and Assignment}
\label{sec:channelAA}

\begin{table*}
\caption{Using genetic algorithms for \textit{\underline{channel allocation/ assignment}} in wireless networks}
\centering

\begin{tabular}{p{2.5cm} p{1cm}p{3cm}p{10cm}}
\toprule
\textbf{\emph{Reference}}    &  \textbf{\emph{No. of \newline objectives}} &  \textbf{\emph{Techniques used}}  &  \textbf{\emph{Brief summary}} \\
\midrule

\multicolumn{3}{l}{\textbf{\emph{Channel allocation in CRNs}}} \\

Zhao et al. \cite{zhao2009cognitive}  & MOGA & Quantum computation with GA\cite{narayanan1996quantum} &  A hybrid GA implemented to decrease the search space and ease the mapping process between the channel matrix and qubit chromosomes.\\

Bhattacharjee et al. \cite{bhattacharjee2011channel}  & MOGA & Comparison of GA with interference graphs &  Channel assignment optimization for single cell cognitive network using GA.\\

Friend et al. \cite{friend2008architecture}  & SOGA & Distributed Island GA &  An Island GA, solves the channel allocation through dynamic spectrum access in cognitive radio networks.\\

ElNainay et al. \cite{elnainay2009channel}  & SOGA & Distributed Island GA &  A localized island GA cooperate with other cognitive network nodes to utilize the available spectrum opportunistically and efficiently through channel allocation.\\

Ali et al. \cite{ali2012genetic} & MOGA & Luby transform codes and GAs  & A GA assisted resource management scheme is presented for reliable multimedia delivery.\\


\\
\multicolumn{3}{l}{\textbf{\emph{Channel allocation in cellular networks}}} \\

Zhenhua et al. \cite{zhenhua2010modified}  &  MOGA & Immuned GA with elitism &  A modified immune GA, integrating immune operators, and minimum separation encoding to obtain the conflict free channel assignment in cellular radio networks.\\

Pinagapany and Kulkarve \cite{pinagapany2008solving}  & SOGA & Parallel GA with decimal encoding  &  A GA based scheme used to solve both fixed and dynamic channel assignment problems in multi-cell cellular radio network.\\ 

Jose et al. \cite{jose2007new}  &  SOGA & GA with Elitism and fitness entropy &  Diversity guided Micro GA in cellular radios to optimize frequency/channel reuse in fixed channel assignment problem.\\

Lima et al. \cite{lima2007adaptive}  &  MOGA  & GA with immigrants and memory schemes &  Two adaptive GAs for locking and switching channel, solve the dynamic spectrum access problem through distribution of channels among cell in cellular networks.\\ 

\\
\multicolumn{3}{l}{\textbf{\emph{Channel allocation for broadband wireless networks}}} \\
Wong and Wassel \cite{wong2002dynamic}  & SOGA & GA based dynamic channel allocation &  GA based technique incorporated channel segregation to maximize the throughput in broadband fixed wireless access networks.\\

Fang et al. \cite{fang2006gop}  & SOGA & GAs with video coding scheme &  Evolutionary technique alongwith Reed-Solomon coding and SNR matrix allocate channel rates for scalable video transmission.\\

\bottomrule
\end{tabular}
\label{tab:channelalloc}
\end{table*}

With the rapid proliferation of wireless and cellular networks, and the problem of limited frequency spectrum, the problem of channel assignment and reuse has become very important. The recent advancement in wireless technology has enabled many high-speed data services leading to an upsurge in number of wireless users. Since the frequency spectrum is a limited resource, there is a great emphasis on the reuse and sharing of frequency amongst the ever-increasing user base. The goal of channel allocation is to maximize frequency reuse by minimizing the number of channels required to cover all the network cells. 

There has been a lot of GA-based work done in the area of channel allocation and assignment in wireless networks (a summary of which is provided in Table \ref{tab:channelalloc}). We discuss some important GA-based channel allocation/ assignment works below.

Broadly speaking, there are three main constraints that channel assignment algorithms in wireless networks must consider: \emph{(i) co-channel interference constraints}---which requires that the same frequency or channel cannot be assigned to two different radios simultaneously; \emph{(ii) adjacent channel interference constraints}---which requires that adjacent frequencies not be assigned to adjacent channels simultaneously; and, \emph{(iii) co-site interference constraints}---that requires minimum spacing between a pair of frequencies assigned to a channel must have minimum spacing. 

Kassotakis et al. have proposed an approach based on hybrid genetic approach---that combines genetic algorithms with local improvement operator---for channel reuse in multiple access telecommunication networks \cite{kassotakis2000hybrid}. The proposed scheme aims to establish the maximal number of connections while minimizing the number of isochronous channels. Wong and Wassel \cite{wong2002dynamic} investigated the channel allocation for a `\textit{broadband fixed wireless access}' (BFWA) networks. A channel allocation matrix keeps track of channel parameters for conflict free assignments. It is observed that as the network grows, centralized scheme would offer unacceptable signaling overhead. GA is used for distributed channel where each access point (AP) transmits the interference power of the entire available channel to a control unit. Control unit employs GAs and ranks the chromosomes of channel priorities according to fitness value of interference. Crossover and mutation rate are also set to promote the best channels and explore the new channels. Experimental Statistics have revealed that better SNR gain (more than 14.5 dB) and throughput is achieved with GAs in comparison with the other two dynamic channel assignment methods: \emph{(i)} least interference \cite{cheng1999wireless}, and \emph{(ii)} channel segregation \cite{akaiwa1993channel}. 

 A study on GAs in channel assignment in cellular networks is performed by Kim et al. \cite{kim1996channel}. This paper reports that neural networks are more likely to get trapped on a local optima since a neural network's output critically depends on the the initial values. In the case of GAs, they possess crossover and mutation as their key strength to escape local optimas. GAs, therefore, have far better chances to overcome the channel allocation problems such as \emph{(i) co-channel interference} \emph{(ii) adjacent channel interference} and, \emph{(iii) co-site interference}. GAs are applied on clusters in cellular networks and it is showed that if the number of cells in one cluster are increased, the convergence rate is decreased leading to an increase in the number of generations due to the reuse of the same frequency among different cells. GA comes handy in keeping the cluster size optimum for efficient frequency distribution among cells.

Ngo and Li \cite{ngo1998fixed} presented a modified GA to contribute in conflict free channel allocation and spectrum efficient assignment in cellular radio networks. There are generally two kinds of channel assignments: \emph{(i) fixed channel assignment (FCA)}---where channel slots are pre-allocated to cells, and \emph{(ii) dynamic channel assignment (DCA)}---where channels are assigned upon request. FCA generally outperforms DCA under heavy traffic load. This algorithm, called the genetic-fix algorithm, generates and manipulates individuals with fixed size, and hence greatly reduces the search space. Furthermore, using the minimum-separation encoding scheme, the required number of bits for representing solutions is substantially reduced. 

\vspace{2mm}
\subsubsection{Localization}
\label{sec:Localization}

\begin{table*}
\caption{Using genetic algorithms for \textit{\underline{localization}} in wireless networks}
\centering
\begin{tabular}{p{1.8cm} p{2.7cm} p{1.3cm} p{2.5cm} p{7.5cm}}
\toprule
\textbf{\emph{Reference}}  &  \textbf{\emph{Goal}}  &  \textbf{\emph{No. of \newline objectives}}  &  \textbf{\emph{Technique used}}  &  \textbf{\emph{Brief Summary}} \\
\midrule

\multicolumn{3}{l}{\textbf{\emph{}}} \\

Zhang et al. \cite{zhang2008localization}  &  Node localization with random distributions & MOGA & GA with simulated annealing &  A GA approach for centralized architecture to optimize the localization in WSN with least mean localization error.\\

Yun et al. \cite{yun2008centroid}  &  Centroid localization  &  SOGA & GA with fuzzy logic &  A hybrid GA with fuzzy membership is developed for edge weights to localize the nodes using received signal strength indicator (RSSI) information.\\

Tam et al. \cite{tam2006using}  &  Localization with ad-hoc positioning system (APS)  & SOGAs & MicroGA as post-optimizer with APS &  A single objective MicroGA, using smaller population worked out best to improve the localization accuracy in APS for both isotropic and an-isotropic topologies.\\

Nan et al. \cite{nan2007estimation}  &  Sensor nodes localization   & SOGA & GA with elitism and partial map crossover (PMX)  &  A real-coded version of GA utilizes roulette wheel selection with Integer coding, thus, helping in achieving precise node localization in sensor networks. \\

Yun et al. \cite{yun2009soft}  &  Localization with RSSI  & SOGA & GA and fuzzy logic system &  Calculation of edge weight with reference of anchor nodes, followed by modeling through fuzzy logic functions and optimization by GA in the end.\\

\bottomrule
\end{tabular}
\label{tab:Localization}
\end{table*}

\begin{table*}
\caption{Using genetic algorithms for \textit{\underline{wireless AP placement optimization}}}
\centering
\begin{tabular}{p{1.8cm}  p{1.8cm} p{4cm} p{9.3cm}}
\toprule
\textbf{\emph{Reference}}  &   \textbf{\emph{No. of Objectives}}  &  \textbf{\emph{Technique used}} & \textbf{\emph{Brief Summary}} \\
\midrule

\\
\multicolumn{3}{l}{\textbf{\emph{Wireless Mesh Networks}}} \\
Smadi et al. \cite{smadi2009free}   & Bi-Objective GA & GA with ILP &  A Joint clustering and gateway problem is addressed by GAs for placement of hybrid free space opticals (FSO) to supplement the gateway capacity in WMNs\\

Xhafa et al.\cite{xhafa2010genetic}  & Bi-objective GA & GA on grid area with rectangular mutation and crossover &  Proposed GAs to optimize the connectivity and user coverage area of mesh network through optimal placement of nodes in the area.\\

\\
\multicolumn{3}{l}{\textbf{\emph{Mobile Ad-hoc Networks}}} \\
Kusyk et al. \cite{kusyk2011self}  & MOGA & Distributed forced GA with evolutionary game theory &  A GA alongwith traditional game theory, running at each node in autonomous network, helps nodes deciding their mobility and organization.\\

Sahin et al. \cite{sahin2008uniform}  & SOGA &  Distributed forced GA &  A Distributed GA adjust speed, location, and direction of mobile nodes over a geographical terrain for military applications.\\

Hoffman et al. \cite{hoffmann2011optimization}  &  SOGA & GA with TDMA table and time slot exchanges.  &  A GA approach to gateway allocation in aeronautical networks and TDMA resource scheduling to minimize the latency.\\

\\
\multicolumn{3}{l}{\textbf{\emph{Wireless Sensor Networks}}} \\
Youssef and Younis \cite{youssef2007intelligent}  & SOGA & GAs on pre-calculation generic routing table  &  A GA places the WSN nodes for disjoint clustering and minimizes the number of hops between sensor and nodes to have efficient intra-cluster communication. \\

Tripathi et al. \cite{tripathi2011wireless}  & Bi-objective & GA and Genetic Programming (GP) &  GP improves the deployment, while GA optimizes the node placement for maximum coverage in WSN.\\

Pandey et al. \cite{pandey2007hybrid}  & MOGA & GA with greedy algorithm and `binary integer linear programming' (BILP) &  A hybrid GAs along with greedy algorithm optimizes the node placements for traffic balancing, in two tier hierarchical heterogeneous WSNs.\\ 

Xu and Yao \cite{xu2006ga}   & SOGA & GA with integer coding and order mapped crossover (OMX) &  A GA approach worked on optimal placement of WSN nodes in grid area full of hindrances and obstacles to match in real-time environment.\\ 

\bottomrule
\end{tabular}
\label{tab:AP_placement}
\end{table*}

The task of determining the location of a wireless node is known as localization. Localization is important in many different settings (such as military monitoring services and disaster management) for providing location-aware services. GAs have been used to assist the localization process in various wireless settings. In this section, we will provide an overview of such work. A tabulated summary of some sample works are presented in Table \ref{tab:Localization}. 

A simple way to localize the nodes is to deploy GPS, but installing GPS in thousands of closely placed nodes is impractical. A better approach is to employ GPS in a few nodes, called anchor nodes, and apply techniques to locate all the other nodes relatively, with reference to multiple anchor nodes. Tam et al. \cite{tam2006using} proposed localization technique in which each node finds its position relative to three anchor nodes, thus, enabling each node to locate itself through triangulation. This information is then broadcasted to base station where a centralized GA works as a post-optimizer, thus, leading to precise network graph with each node's location known. Zhang et al. presented the centralized GA with \emph{simulated annealing}\cite{zhang2008localization} to minimize the error in position estimation of nodes in WSNs. This approach is also based on anchor nodes localization scheme. Simulated annealing is local search technique and it avoids pre-mature convergence, by mixing inferior and sub-optimal solutions with the optimal ones. In the approach of \emph{centroid localization} \cite{yun2008centroid}, proposed in \cite{yun2008centroid}, GAs and fuzzy logics are used by each node to perform self-localization by estimating the edge weights based on the received signal strength (RSS).

\begin{table*}
\caption{Using genetic algorithms for \emph{\underline{general applications}} in wireless networks}
\centering
\begin{tabular}{p{2cm} p{0.8cm} p{1.5cm} p{3.7cm} p{8cm}}
\toprule
\textbf{\emph{Reference}}  &  \textbf{\emph{Network Type}}  &  \textbf{\emph{No. of \newline objectives}}  &  \textbf{\emph{Technique used}} &  \textbf{\emph{Brief summary}} \\
\midrule

\multicolumn{3}{l}{\textbf{\emph{Area Coverage}}} \\

Xhafa et al. \cite{xhafa2010genetic}  & WMNs  & Bi-objective GA & Rectangular mutation \&  crossover &  A GA approach is used to maximize the `giant component' and coverage area in WMNs while using multiple distribution strategies of client nodes.\\

Alba et al. \cite{alba2007cellular}  &  MANET &  MOGA &  Broadcasting updates from base station running GA & A hybrid broadcasting protocol using evolutionary algorithms. A MOGA approach is used to optimize metropolitan-scope MANET coverage and usage.\\

Hu et al. \cite{hu2010hybrid}  & WSN & SOGA & GAs with schedule transition operations &  Hybrid GA with scheduled transition and forward encoding scheme finds the maximum number of disjoint cover sets to enhance the disjoint covered sets in areas. \\

Sahin et al. \cite{Sahin:2008:GAS:1389095.1389318}  & MANET & MOGA & A Distributed forced GA &  A forced GA is proposed letting the nodes decide their movement, direction and speed through molecular force equilibrium to get maximum area coverage\\

Lai et al. \cite{lai2007effective}  &  WSN  & SOGA & Integer coding &  GA Based scheme with one-to-one mapping of genes and sensors to find maximum disjoint sensor covers to prolong the network life span. \\

Quintao et al. \cite{quintao2005evolutionary}  &  WSN  & SOGA & GAs coupled with mathematical modeling and ILP &  GA based heuristic approach with integer linear programming (ILP) and mathematical formulation to solve dynamic network coverage problem. \\

\\
\multicolumn{3}{l}{\textbf{\emph{Network Planning and design}}} \\
Bhondekar et al. \cite{bhondekar2009genetic}  &  WSN  & MOGA & GA model for grid deployment &  A GA decides the transmission ranges and the active/ sleeping nodes in WSN topology while accounting for connectivity, energy, and other application specific parameters.\\

Jourdan et al. \cite{jourdan2004layout}  & WSN & MOGA & GA for non-linear objectives & A MOGA approach design the Pareto optimal layouts for WSN nodes depending upon the sensing and communication range.\\

Pries et al. \cite{pries2009wireless}  &  WMNs  & SOGA & Tree coding &  GA operators like subtree crossover and cell crossover, respectively, are employed in solving small and large-scale WMN gateway optimization problems.\\

Ayyadurai et al. \cite{ayyadurai2011multihop}  & CN & SOGA & Single cell optimization  & An addaptive GA for dynamic system reconfiguration optimizes the resource allocation in cellular networks.\\

Ghosh et al.\cite{ghosh2005gama}  &  WMN  & SOGA & Binary encoding of nodes &  A dynamic mesh topology design, using GAs, while maintaining low network cost and path redundancy at base station for reliability.  \\

Sheen et al. \cite{sheen2010downlink}  &  CN  & MOGA & Joint optimization & A GA for maximizing joint multi-cell CN system efficiency, through optimizing relay stations (RS) placement, frequency reuse pattern, and maximizing throughput to enhance the end user experience.\\

Chiaraviglio et al. \cite{chiaraviglio2012energy}  & CN  &  MOGA & Hybrid GA with centralized sorting of base stations &  GA based switch off strategies for enabling sleep modes in base stations to design energy efficient cellular networks.\\

\\
\multicolumn{3}{l}{\textbf{\emph{Cluster Optimization}}} \\

Zeng et al. \cite{zeng2008load}  &  WMN  & MOGA & Greedy algorithm with GAs & A hybrid GA meets the QoS and load balancing constraints separately by constructing the graph and dividing the WMNs in disjoint clusters. \\

Hussain et al. \cite{hussain2007genetic}   &  WMN  & SOGA & Cluster optimization through distributed broadcasting &  GAs augmented the network life when compared with LEACH \cite{heinzelman2000energy} by employing energy efficient data dissemination and optimizing the cluster distances.\\

Zhang et al. \cite{zhang2008optimal}  &  WSN  & SOGA &  Simulated annealing (SA)\cite{stumpf1994enhanced}  &  A GA approach provided the energy efficient WSN topology by improving the data aggregation rate and minimizing the intra-cluster distances.\\

Huruiala et al. \cite{huruiala2010hierarchical}  &  WSN  & SOGA & Centralized GA &  A GA running on BS coordinate with nodes to minimize latency and power consumption for routing under tight constraints.\\

\bottomrule
\end{tabular}
\label{tab:generalApplications}
\end{table*}

\vspace{2mm}
\subsubsection{WLAN AP placement}
\label{sec:APplacement}

Since the arrival of IEEE 802.11, WLAN services are widely adopted with the predominant WLAN setting being infrastructure mode (in which the topologies around APs). Optimizing the placement of WLAN APs plays a substantial role in maximizing the throughput and data rate of WLANs. In this section, we will summarize the various proposals for using GAs in WLAN AP placement. A tabulated summary is presented in Table \ref{tab:AP_placement}.

A multi-objective GA (MOGA) approach for WLAN AP placement (with the objectives of minimizing the number of APs and maximizing SNR) has been described by researchers in \cite{maksuriwong2003wireless}. GA-provided solutions have lesser variance implying that a larger number of area points within the network would have average SNR. Statistical analysis reveals that MOGA approach generate solutions with a high SNR data profile.

Optimizing the radio coverage inside buildings is as important as outdoors radio optimization. WLANs are mostly used inside homes and offices to provide mobile and Internet coverage to subscribers. Nagy and Farkas \cite{nagy2000indoor} have studied the problem of optimizing the placement of indoor Base Stations (BS) while taking in account the effects of floors, ceiling, and walls on signal propagation and path losses. GA used the microwave propagation model to account for the effect of influence of brick, concrete, wood, furniture, and glass obstacles. Different weighting coefficients were assigned to each obstacle and parallel heuristic approach of GA is used to optimize the coverage results within indoor environment.

Another important application of GAs is in the optimization of AP placement in cellular and mobile networks. Since finding the set of coverage areas whose sum is the maximum covered area is an NP-hard problem, GAs come handy. Lieska et al. \cite{lieska1998radio} have explored in their study the combination of simulation and evolutionary techniques. Radio coverage areas are obtained by the series of BS with received power greater than -60dbm. Algorithm works by selecting the chromosomes based on required BS. The chromosomes are than evaluated against constraints of power and SNR; afterwards, the healthy chromosomes are allowed to reproduce leading to global convergence of the entire network. This fitness function can be modified to include other constraints based on network requirement. A major advantage of a GA, therefore, is its ability to adapt to multiple situations and problems.

\vspace{2mm}  
\subsubsection{General Applications}
\label{sec:GeneralApplications}

A survey for solving computationally difficult problem, through GAs, in wireless networks is presented by Das et al. \cite{das2006solving} in his book ``Handbook of Bio-Inspired Algorithm and Applications''. It has been shown that GAs have the potential to solve any optimization problem provided that proper modeling and chromosomes encoding has been performed. In this section, we will cover some of the other miscellaneous applications of GAs in wireless networking (that we have not covered in previous sections). A tabulated summary of these applications are presented in Table \ref{tab:generalApplications}. We describe some of these applications here; some other applications are discussed when we cover network-configuration based categorization of GA applications in Section \ref{sec:networkbased}).

\vspace{2mm}
GAs have been proposed for \textit{energy efficient planning and management of cellular networks} by enabling sleep modes in base stations \cite{chiaraviglio2012energy}. Johnson and Samii \cite{johnson1995genetic} have proposed a GA based approach to find the best node distribution with optimized antenna pattern and SNR. The authors have employed simple binary coding scheme for a pool of randomly generated chromosomes for different nodes distribution and evaluated their fitness against the fitness function (in which $N$ is the total number of nodes):

\begin{equation}
 Fitness=max-\frac{1}{N} \sum_{i} Path\_length_{i}
\end{equation}

The winner chromosomes (having minimum cost of path length) are more probable to be inherited in the next generation. This approach is also applied for random encoding of antenna pattern and transmission power.

\vspace{2mm}
\textit{Churn prediction and forecasting} have application in almost all the real world phenomenon including medical, finance, products sale, telecommunications and network. Forecasting the future sale and growth in turbulent markets like telecommunications is always a big issue for product managers. Even approximate insights can yield large profit return in the long run. Venkatesan and Kumar studied GAs for forecasting \cite{venkatesan2002genetic} the magnitude of future sales, time period to peak sales and the value of sales during peak time. Considering the significant advantages of accurate forecast of product lifecycle, GAs are used to estimate the diffusion model based on previous data points. GAs are fed with several critical factors like price, competitive intensity and network effects in obtaining better predictions during growth phase. Since GAs are inherently parallel, so, different scenarios can be taken care of while making predictions. Study shows that predictions by GAs are robust and accurate in predicting the future growth of a market. Pendharkar \cite{pendharkar2009genetic} presented the hybrid GA with neural networks to forecast the churn in cellular network services. The hybrid algorithms optimize the objectives after learning and assigning weights to certain parameters. Results obtained from GA method are shown to be more accurate than statistical models but are computationally expensive. 

\subsection{\textbf{Network Type Based Classification}}
\label{sec:networkbased}

In this section, we will describe the applications of GAs in wireless networking categorized according to the type of wireless network. In particular, we will present GA applications in the networking configurations of wireless sensor networks (WSNs), wireless mesh networks (WMNs), cognitive radio networks (CRNs),  cellular networks (CNs), and software-defined wireless networks (SWNs).

\vspace{2mm}  
\subsubsection{Wireless Sensor Networks (WSN)}

WSNs have emerged as an exciting technology that can be used to monitor the physical world. WSN is composed of self-organizing, miniaturized sensors nodes, running on batteries, and communicating over wireless links. The major challenge in WSNs is to provide area coverage with minimum WSN nodes while preserving the resources and energy consumption.

Jia et al. \cite{jia2009energy} addressed this problem through MOGA for energy efficient coverage control in WSNs. The algorithm provides a mechanism to cover the maximum deployed area with minimum number of WSN nodes. Since connectivity and coverage are both dependent on each other so this problem is addressed by MOGA approach by achieving the Pareto optimal front. This is a non-dominated problem so it comprises of multiple solution individuals that are evaluated against two fitness function coverage area and sensor used rate. The MOGA approach applied performs well in maximizing the area covered with least computation. Another advantage is that it avoids partial solutions in whole search space and reset sensor nodes once in whole network simulation.
 
Frentinos and Tsiligiridis \cite{ferentinos2007adaptive} studied an adaptive WSN design based on MOGA. Individual with random placement of nodes are utilized as population and GAs decides which sensors would be active and which would act as clusterhead. Random designs are encoded and fed to GAs and the winner chromosomes are going to decide the network infrastructure including clusterheads and active or passive nodes along with signal strengths. From evolution of characteristics, it was shown that generally, it is more favorable to have large number of sensor nodes with low energy rather than having lesser number of sensors with high energy consumption. This application of adaptive GA in WSN can lead to increase in network life span while maintaining network properties close to optimal value as well as it helps in adapting sophisticated decisions about operating modes (clusterheads, active nodes with optimal signal range).

Extending the life span of any WSN network is active area of research. Different schemes have been proposed like centralized monitoring GA for keeping active and passive nodes in network. A similar approach has been discussed by Jin et al.  \cite{jin2003sensor} to cut the energy consumption in WSNs. A GA is used for independent cluster formation to reduce the communication distances thus bringing the route cost down. Experimental results showed that by keeping the number of cluster to 10\% of total nodes, results in 80\% reduction in path cost and communication distance as compared to random distribution of WSNs. In another work, Sengupta et al. \cite{sengupta2012evolutionary}  have proposed an evolutionary multiobjective sleep-scheduling scheme for differentiated coverage in WSNs.

Zhang et al.\cite{zhang2008genetic} used GA to extend the work of Niculescu and Nath on network localization  in WSNs \cite{niculescu2003ad}. The central idea is to employ GPS enabled nodes, called anchor nodes, in the network and to have all the other network nodes estimate their position through 1-hop distances to the anchor nodes. The locations are then flooded by each node to the clusterhead which uses the location information to construct the graph of the entire network. By using standard GA operators like mutation and crossover, the proposed GA based approach was able to better estimate locations resulting in lesser mean position error than other algorithms in literature such as the `\textit{semi-definite programming with gradient search localization}' (SDPL) \cite{liang2004gradient} and the `\textit{simulated annealing based localization}' (SAL)\cite{kannan2005simulated}.


In another work, Hussain et al. \cite{hussain2007genetic} have studied and simulated GAs for centralized hierarchical WSNs. The proposed scheme comprises a  hierarchy of layers with the BS being at the top layer and the clusterheads and the sensor nodes occupying the bottom layer. GAs are used to construct clusters with a clusterheads and other nodes. The BS layer keeps track of the entire cluster through their clusterheads. A GA approach minimizes the average cluster distances while simultaneously increasing the total number of transmissions. After configuring sensor nodes in optimized formation, the data transmission phase follows.

\subsubsection{Wireless Mesh Networks (WMNs)}

WMNs provide a popular mechanism for the provisioning of cost-effective wireless broadband connectivity. WMNs are composed of (potentially mobile) mesh clients and (typically fixed) mesh routers that connect over a relatively static multihop wireless network \cite{chou2006low}. To address the issue of efficient placement of mesh routers, Xhafa et al. \cite{xhafa2010genetic} employed GAs to optimize the user coverage area and connectivity of mesh network through optimal placement of nodes in an area. Chromosomes are designed in rectangular grids having information of both clients and mesh router nodes. Crossover is implemented to preserve the chromosomes traits by randomly exchanging parent and child router nodes from one cell to another in grid area, thereby promoting high scoring chromosomes. Mutation is then utilized to randomly migrate a router to another grid and recalculating the fitness. Various models can be adopted for distributing client nodes such as the normal, uniform, and exponential distribution. GAs are excellent in computing the mesh router placements and always succeed in achieving the connectivity, irrespective of client distribution, however, area coverage is subjected to client nodes in network area.

\vspace{2mm} 
\subsubsection{Cognitive Radio Networks (CRNs)}

Cognitive radio networks (CRNs) offer a potential solution for improving the spectrum utilization in the world of limited spectrum. In dynamic spectrum access (DSA) based CRNs, the secondary users of a network aim to utilize the spectrum opportunistically to offer QoS for secondary users without causing any interference to licensed primary users. Such DSA based CRNs utilize cognitive radios (CRs) that incorporate various AI techniques for intelligent decision making \cite{qadir2013artificial}. The main benefits of applying GA in CRNs are that: \emph{(i)} it is conceptually easy to understand; \emph{(ii)} it is inherently amenable to parallel solutions and therefore can be easily distributed; and that \emph{(iii)} GAs lend support to multi-objective optimization and performs well in noisy unknown environments. 

An early application of GA techniques to CRNs is documented in a paper authored by Rondeau et al. \cite{rondeau2004cognitive}. This paper presented the adaptation mechanism of a cognitive engine, implemented by the authors, that used GAs to evolve a radio's parameters to make them best for the user's current needs. A `\textit{wireless system genetic algorithm}' (WSGA) is also proposed to realize adaptive waveform control and cross-layer optimization \cite{rondeau2004cognitive}. GAs have also been used for building distributed \emph{parallel} solutions, using the technique of \emph{island genetic algorithms}, to mitigate the channel assignment problem in CRNs \cite{friend2008architecture}. An island GA divides the overall population into subpopulations known as islands, which then interact by migrating individuals to the other island. More details of parallel genetic algorithms, and of general parallel meta-heuristics, can be found in the survey \cite{alba1999survey} and \cite{alba2005parallel}, respectively.

\vspace{2mm}
\paragraph{MOGAs for optimization of CRNs}

Multi-objective optimization aims at estimating a true Pareto front optimization in wireless medium depending upon the external environment parameters. These radio environment parameters serves as genes and are encoded in binary strings, which form chromosomes. These chromosomes are evaluated against fitness function regularly in successive generations. After realizing sufficient fitness or completing predefined number of generation, these set of transmission parameters are sent for updating new radio configuration. Due to variation in radio channel characteristics and spectrum availability, a cognitive radio has to maintain time varying QoS while optimizing different parameters such as BER, throughput, power consumption, and interference. 

%
%

MOGAs have been used extensively for optimization of cognitive radio networks (CRNs) \cite{fette2009cognitive}. The foremost aim of cognitive engine (CE) is optimizing the radio parameters for best QoS and secondary goal is to observe and learn for adaptation. To achieve these goals cognitive engines have \emph{(i)} cognitive system module (CSM)---for modeling and learning \emph{(ii)} Wireless System Genetic Algorithm (WSGA)---for actions \emph{(iii)} Wireless Channel Genetic Algorithm (WCGA)---for hardware implementation of radio parameters. These parameter and radio traits called genes represent chromosomes, are modulation, frequency, spreading, filtering, and FEC coding, etc. Different objectives are optimized based on relative weights assigned to them using multi-objective approach. These objectives include optimizing BER, power consumption, interference and MOGA approach is applied on population and chromosomes, which performs best in most of the objective, are chosen as new set of radio parameters.

\paragraph*{Case Base Decision Theory (CBDT) for Initialization of GAs}

Since the secondary objective of CR is to adapt and evolve with changing environmental parameters by updating radio configurations. So a CR system must be provided with a memory database that continuously monitors and learns GA outcomes in real-time optimization. As the system matures, with the passage of time, each new optimization would take less time to achieve global optima. More thoroughly speaking, CE keeps track of previous objectives and the winner chromosomes. So whenever a new objective or case is encountered, it is compared to previous cases in memory and chromosomes of the case/objective having maximum similarity and utility are used to partially initialize the GA population for current objective. Rest of the population is randomly generated to not only maintain diversity but also to help in achieving optima in lesser number of iterations. This adaptation scheme is carried out through algebraic manipulations with the independently constructed channel availability matrix, the interference matrix, and the channel reward matrix (which are each kept up  to date about the external radio environment conditions) \cite{ye2010genetic}.

Rieser \cite{rieser2004biologically} presented a biologically inspired cognitive radio model, using modern AI techniques. This model is based on adaptive cognitive engine that learns the radio environment, optimizes the radio parameters, and presents the best possible experience to secondary user. He built his foundation of GA based technique on original Mitola \cite{mitola2006cognitive} proposed technique of software radios. He gave insights on all the popular AI Algorithms and how ineffective they are in evolving as compared to GAs, when faced with unanticipated radio channel environment. He applied GA to hidden Markov model for error modeling in uncertainties in radio channel environment. Wireless channel GA has its foundation on these radio environmental modeling.

\begin{figure}[ht!]
\centering
\includegraphics[width=.5\textwidth]{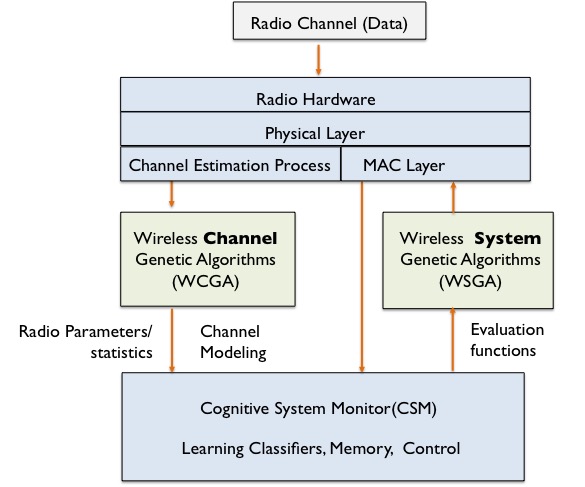}
\caption{Cognitive Engine Architecture with WCGA, CSM, and WSGA \cite{fette2009cognitive}}
\label{fig:CEarchitecture}
\end{figure}

The biologically inspired cognitive engine serves and evolves well in unanticipated radio channel environment, which make it usable for military purpose and emergency situation. This whole cognitive architecture works in multiple steps. Firstly, the engine gathers the channel information which is then subjected to machine learning techniques such as learning classifiers. After learning and modeling, these information are taken up to alter the radio configurations for tweaking appropriate settings \cite{fette2009cognitive}. Wireless system GA treats the radio as the analogue of a biological creature, and makes optimized discoveries to keep the appropriate balance of radio parameters. The cognitive engine architecture, with WCGA, CSM, and WSGA, is displayed in Figure \ref{fig:CEarchitecture}. We will only briefly explain below the main infrastructural components of the cognitive engine developed by Rieser. 

\vspace{2mm}  
  
\begin{enumerate}

\item \emph{Wireless Channel Genetic Algorithm (WCGA)}: WCGA takes the information about radio channel through physical and MAC layer and constructs a statistical model based on external channel environment. This information is then forwarded to CSM.
  
\item \emph{Cognitive System monitor (CSM)}: CSM incorporates the capability of learning and adaptation. CSM after learning and generating appropriate action goals instructs WSGA \cite{fette2009cognitive} for actions leading to different radio configurations.
  
\item \emph{Wireless System Genetic Algorithm (WSGA)}: WSGA receives the channel model by CSM with fitness function and initial chromosomes. WSGA performs all the GA computation through mutation, selection and crossover, thereby helping to select the best chromosomes. These chromosomes are, in fact, different radio configurations, which are then used for implementing new radio settings.

\end{enumerate}

\vspace{2mm}
\paragraph{Cognitive radio (CR) testbeds}

Rieser \cite{rieser2004biologically} developed a biologically inspired cognitive radio CR testbed. A number of cognitive theories are developed and evaluated in parallel through the use of GA operators. This developed testbed consists of multilayered cognitive engine inspired by human brain evolution. The architecture is shown in figure. It provides multilayered scaled functionality with sensing radio parameters at both MAC and physical layers. It uses three algorithms that can be co-located in same radio or distributed to multiple radios. This cross-layered cognitive engine architecture is shown in Figure \ref{fig:CEarchitecture}. Another GA-based software testbed for cognitive engine (CE) is proposed by Kim et al. \cite{kim2008cognitive}. A two-step GA is used to avoid problem of optimizing the biased dominant parameters. To implement this, GA based iteration is run for channel selection and transmitter parameters selection separately. Simulation of this testbed demonstrates that bandwidth and transmit power converge to optimal values with in few iterations of GAs.
 

\vspace{2mm} 
\subsubsection{Cellular Networks (CNs)}

Next generation wireless networks have to cope with the massive increase in multimedia service along with cellular data. This motivates the development of self-consolidating networks that are capable of making intelligent decision to offer the best user quality of experience (QoE). Ayyadurai et al. \cite{ayyadurai2011multihop} presented an adaptive GA to optimize the cellular networks spectral efficiency through dynamic allocation of resources, between the relay stations and the base stations. Population adjustment came along the way to improve the search capabilities leading to quick optimum. Chiaraviglio et al. \cite{chiaraviglio2012energy} presented the energy efficient deployment of cellular networks by leveraging GA to minimize the power consumption and number of transmitter in cells. This involves employing sleep modes, where, a base station decides which nodes should be powered on, depending on number of active users in given cell.

A conflict free channel assignment is an NP-hard problem. A powerful GA incorporating immune operators including immune clone, immune vaccination, and immune selection has been proposed by Zhenhua et al. \cite{zhenhua2010modified} to mitigate the interference constraints and obtain a conflict free channel assignment. Vaccination involves changing some bits of a gene to have an improved fitness value. Elitism keeps the population diversity and minimum separation encoding helps in keeping a frequency distance among individual channels. A similar approach has been studied by Pinagapany and Kulkarve \cite{pinagapany2008solving} for solving both dynamic and fixed channel assignments in cellular networks. Channel compatibility matrix is constructed first to record the violation of individuals. After that, cell and bandwidth information are encoded in individual chromosomes, which are then subjected to crossover and mutation. Jose et al. \cite{jose2007new} offered low computational micro GA with elitism strategy to solve the fixed channel assignment (FCA) problem. The proposed algorithm spotlighted how to keep diversity by varying mutation probability to avoid local optima. 

In cellular networks, it is desirable to minimize the call drop and call blocking probability. Mobile cellular networks are often evaluated against the criteria of dropping probability of current calls and the blocking probability of new calls. Lima et al. \cite{lima2007adaptive} studied two adaptive GAs for locking and switching channel to solve the dynamic spectrum access problem in cellular networks. The proposed fitness function evaluates the capacity of each channel for new calls and also keep tracks of existing calls. The algorithm takes care of randomness by allowing random immigrants and parameter adaptation to keep exploring new option for each channel.

\vspace{2mm} 
\subsubsection{Software Defined Wireless Networks (SWNs)}

In recent times, software defined networking (SDN) has been proposed as a new architecture for conveniently operating and managing diverse wireless networks. SDN is generally understood to entail \textit{(i)} the separation of the control and data plane; (\textit{ii)} the control of multiple data planes by a single controller; and the \textit{iii)} development of new sophisticated control abstractions \cite{qadir2014prog}. Wireless networks using SDN principles are referred to as software defined wireless networks (SWNs). Although this networking configuration is relatively new, the use of GA has been proposed in SWNs. For example, Bojic et al. have proposed a small-cell mobile backhaul resource manager, based on GA, that interworks with software defined management \cite{bojic2013advanced}. 

\subsection{\textbf{Hybridization: GA With Other Techniques}}
\label{sec:Hybrid}

GAs is a useful framework that can be profitably used in a wide variety of settings. It is also possible to use GAs with other adaptive algorithmic  approaches such as game theory, reinforcement learning (RL), graph theory, fuzzy logic based and quantum theory based systems. Such a symbioses is not at all antithetical from the theme underlying the GA framework which encourages pluralism in preference to relying on any one solution; it is often the case that we can perform better by combining the best parts of different solutions as building blocks. In this section, we will describe hybrid approaches that combine various techniques with the principles of GAs. 

\vspace{2mm}
\subsubsection{Combining GAs with Game Theory}

Game theory is a mathematical decision framework through which we can study and analyze competitive interaction between two or more self-interested rational agents. The field of evolutionary game theory utilizes the concepts of evolutionary computing and genetic algorithms in the backdrop of a game-theoretic background \cite{weibull1997evolutionary}. Evolutionary games have been used in a wide variety of contexts in networks including cooperative sensing \cite{wang2010evolutionary}, multiple-access-control and power control \cite{tembine2010evolutionary}, routing games \cite{altman2007evolutionary}, network selection games \cite{niyato2009dynamics}, congestion control \cite{altman2008evolutionary}, adaptive TCP \cite{anastasopoulos2010tcp}, and for dynamic bandwidth allocation \cite{zhu2010optimal}. The interested reader is referred to a chapter on this topic in a recently published book on game theory in wireless and communication networks \cite{han2012gamech} for a detailed treatment.

\vspace{2mm}
\subsubsection{Combining GAs with Reinforcement Learning}

Reinforcement learning, unlike supervised learning, is a technique where agent has no prior knowledge of transitions and learns along the way through experience and interaction with the environment. We note that evolutionary algorithms like GAs are related to reinforcement learning as they relies on exploration and exploitation \cite{vcrepinvsek2013exploration}. More specifically, GA utilizes the genetic operators of mutation and recombination as procedures for exploration along with selection procedures to exploit known good solutions. Reinforcement learning can be carried out both by searching through the policy space or the value-function space. Genetic algorithms can be thought of as reinforcement learning scheme of the policy space searching variety. Policies are encoded in chromosomes and deterministic procedure is used to evaluate the cumulative fitness of the agent the follows a given policy. According to evolutionary principle of survival of the fittest, only good policies are adopted while  failed or weaker policies are relinquished \cite{moriarty1999evolutionary}. 


\vspace{2mm}
\subsubsection{Combining GAs with Graph Theory}

In order to produce robust dynamic graph-theoretic solutions, graph theory can be combined with GAs in particular, and with evolutionary algorithms more generally. The combination of graph theory with evolutionary algorithms is known as evolutionary graph (EG) theory in literature. Ferreira et al. \cite{ferreira2010performance} have applied EG theory in MANETs with known connectivity pattern for developing efficient practical routing protocols. The dynamic behavior of network is captured through evolving graphs. 
There are various metrics that can be used for determining optimal routes in EG-based models with known connectivity patterns, including the three metrics of \emph{shortest}, \emph{fastest}, and \emph{foremost} proposed in \cite{xuan2003computing}. These metrics determine respectively the minimum number of hops, the minimum time span, and the earliest arrival data. Ferreira et al. \cite{ferreira2010performance} utilized the \textit{foremost journey} metric along with the EG theory. The proposed algorithm can estimate the congested nodes and suggest the parallel paths. In another work, a bio-inspired graph theoretic solution for ensuring energy efficiency in cooperative communication networks has been proposed in \cite{DBLP:journals/corr/GajdukUBK14}. The presented algorithm combine the graph theory with evolutionary techniques to the let the nodes operate in decentralized fashion for intelligent decision making, and do forwarding by selecting energy conserving nodes.


\vspace{2mm}
\subsubsection{Combining GAs with Fuzzy Logic}

Fuzzy logic is a logical system that can entertain logical variables that take continuous values between 0 and 1 instead of discrete values of either 0 or 1 (as in classical digital logic). Fuzzy logic is as an effective method for knowledge representation, control implementation, and cross-layer optimization in wireless networks \cite{baldo2008fuzzy}. Fuzzy logic is well suited for knowledge representation in wireless networks since network conditions in wireless networks are often incompletely and imprecisely known. Fuzzy systems that utilize fuzzy logic are highly suited to problems with imprecise, noisy, and incomplete input information. GA can be used for both optimizing the fuzzy membership function as well as the fuzzy rules that drive the fuzzy control system.  The intersection of hybrid technologies of GA, fuzzy logic, and other AI techniques (most prominently neural networks) is studied in the research theme of `soft computing' \cite{zadeh1994fuzzy}.  The trend of genetic fuzzy systems for soft computing applications is less developed than the trend of neuro-fuzzy systems, especially in the context of industrial control applications. Nonetheless the use of genetic fuzzy techniques is popular for both communication systems \cite{wang2003soft} and wireless networks \cite{yun2009soft}. 


\vspace{2mm}
\subsubsection{Combining GAs with Quantum Theory}

Quantum theory derives its root from quantum mechanics, i.e., mechanics of atoms. A comparative study has been performed in \cite{narayanan1996quantum} between quantum GA and classical GA. Quantum GA proves to be faster and more robust then classical GA for traveling sales person problem. A quantum GA is proposed in \cite{luo2010quantum} to solve the \emph{NP-Complete} QoS metrics problem in WSNs. Quantum GA operate on \emph{qubit} as a smallest unit of information, that serves as a linear superposition of two states, i.e., 1 and 0. A rotation gate is applied to keep the diversity of population in quantum GA. Quantum GA behaves more effectively than traditional GA for larger input set owing to its linear superposition states.

\vspace{2mm}
\subsubsection{Combining GAs with Local Search}

A very common form of a hybrid GA, sometimes referred to as \textit{Memetic algorithms} \cite{moscato2010modern} in literature, is created by combining GAs with local search techniques and to incorporate domain-specific knowledge in the search process \cite{sastry2005genetic}. This can help produce stronger results but at the cost of higher computational costs. The local search operator is applied to each member of the population after each generation. Memetic algorithms have been used in the context of wireless networks for various tasks such as energy-aware topology control \cite{konstantinidis2007energy} and cell assignment to switches \cite{quintero2002memetic}.

\vspace{2mm}
\subsubsection{Combining GAs with Other Metaheuristics}

A popular area of research is combining GAs with heuristic methods like ant colony optimization (ACO), neural networks (NN), and fuzzy logics to solve the dynamic constrained multi-objective optimization. In particular, there has a been a lot of work in applying GAs for adjusting the connection weights and the connectivity itself in NNs \cite{whitley1990genetic}. Notwithstanding the amount of work done, there is a lot of scope for further optimization and improvement. 

\section{Practical Issues and Lessons Learnt}
\label{sec:lessons}

In this section, we will summarize the main insights and lessons learnt through decades of GA research that can guide successful problem solving through GAs. We first describe common pitfalls in Section \ref{sec:pitfalls}, and then describe some guidelines and insights for successful GA implementation in Section \ref{sec:insights}.

\subsection{Common Pitfalls}
\label{sec:pitfalls}

While GAs provide a useful framework for problem-solving in a wide variety of domains, it is not a panacea. In this section, we will highlight some deficiencies of GAs and how it may be incorrectly applied.

\vspace{2mm}
\subsubsection{Genetic Drift or Bias}

In GAs, there is a risk of slow convergence and settling for local minimas, especially with inappropriate choice of model parameters. While GAs can often converge to the global optima in the majority of applications, there are no mathematical guarantees to this effect. In particular, a GA can get get stuck in the vicinity of a suboptimal point if it lacks population diversity, or if has an inappropriately small population. Such a phenomenon is known as `genetic drift' or 'genetic bias'. Certain problems cannot be solved by means of simple genetic operations. This is mainly due to the choice of poor fitness functions that generate poor chromosome blocks. There cannot be an absolute assurance that a GA will be able to find a global optimum---this can especially be a problem when the population has a lot of subjects. 

\vspace{2mm}
\subsubsection{Using GAs for Real-Time Applications}

Like most other AI techniques, GAs are not well suited to real-time application that require guaranteed response times due to its stochastic and heuristic nature. In addition, the response time of GA can have significant variance depending on the various control parameters and population initialization. This potential pitfall should be kept in mind if GAs are used for online learning and optimization in wireless networks. In particular, QoS requirements in wireless networks may put constraints on achieving the optimum decision in limited time forcing GAs to terminate after predefined number of iterations.

\vspace{2mm}
\subsubsection{GA Deception}

It has been reported in literature that certain objective functions, referred to as GA-deceptive functions, can be very difficult to optimize. With such deceptive functions, combination of good building blocks may result in very bad chromosomes, causing the building block theory to fail. 
A potential solution to this problem is using messy genetic algorithms \cite{mitchell1998introduction}.

\subsection{Implementation Insights}
\label{sec:insights}

\vspace{2mm}
\subsubsection{Setting Population Size}

The size of the population in a GA can be a major factor in determining its quality and convergence speed. Generally speaking, larger populations result in better solutions.  A few studies have addressed the problem of determining exactly how large a population should be to ensure a certain quality \cite{goldberg1989sizing}\cite{goldberg1991genetic} \cite{harik1999gambler}. For a moderately complex problem, a population of 50 to 100 chromosomes can serve as a good default value \cite{cox2005fuzzy}. The precise choice of the population size depends on the problem complexity and the number of search variables. While it is typical for the population size to remain fixed from one generation to the next, it can also be made to grow or contract depending on the rate of convergence, and other factors such as the current genetic diversity.  

\vspace{2mm}
\subsubsection{Designing Solution Encoding and Fitness Function}

The efficiency, processing power, and the convergence of GAs depend critically on how the solution is encoded in the form of chromosomes and on the formulation of an appropriate fitness function. This is the first necessary step towards an efficient implementation. There is no one-size-fits-all solution here, and the correct model choice depends on what is most useful for the problem at hand. The encoding of chromosomes should be carefully considered. While a majority of implementations utilize bit representation, due to the ease with which crossover and mutation can be implemented, other representations, utilizing data structures, may be appropriate for the considered problem and should also be considered \cite{sastry2005genetic}. Since there can be numerous encoding and formulations, this step involves considerable ingenuity. Some guidelines on these issues are provided in \cite{grefenstette1986optimization} \cite{de1991analysis}.

\vspace{2mm}
\subsubsection{Setting of Mutation and Crossover Rates}

The optimization of the control variables of genetic algorithms (such as the setting of mutation and crossover rates) plays an important role in the performance of a GA based framework. A GA framework can avoid prematurely settling in a suboptimal rut by emphasizing \emph{exploration} using mutation.  Various solutions have been proposed to the problem of genetic drift/ bias including fitness sharing \cite{fonseca1995multiobjective}, crowding and pre-selection \cite{mahfoud1992crowding}. Bhandari et al. \cite{bhandari1996genetic} showed that the probability of fast convergence increases by seeding the initial population with high scoring chromosomes. As a general rule-of-thumb, crossover and mutation rates of 0.6 and 0.001 have been proposed for large population sizes (of $\sim$100), with corresponding values of 0.9 and 0.01 proposed for small population sizes (of $\sim$30) \cite{grefenstette1986optimization} \cite{de1991analysis}. Sastry et al. \cite{sastry2005genetic} similarly recommend a mutation probability of .05 and a crossover rate of 0.6 with a population size of $\sim$50. A tabulated summary of mutation and crossover rates generally recommended by the GA community and accepted as de facto `standards' are presented in Table \ref{tab:guidelines}.

\begin{table*}
\caption{Summary of GA paramater settings guidelines generally accepted by the GA community}
\centering
\begin{tabular}{p{3cm}cccc}
\toprule
 & \textbf{De Jong and Spears \cite{de1991analysis}} & \textbf{Grefenstette \cite{grefenstette1986optimization}} & \textbf{Sastry et al. \cite{sastry2005genetic}} & \textbf{ Carrol \cite{carroll1996chemical} (Micro GA)} \\
\midrule
    Population Size & 50 & 30 & 50 & 5  \\
    Number of Generations & 1000 & Not Specified & Not Specified & 100 \\
    Mutation Rates  & 0.001  & 0.01  & 0.05  & 0.02; 0.04 \\
    Mutation Types  & bit flip & bit flip  & Not Specified & jump and creep\\
	Crossover Rate  & 0.6 & 0.9  & 0.6 & 0.5\\
    Crossover Type  & 1 or 2 point (typically, 2 point)  & 1 or 2 point  (typically, 2 point) & Not Specified & Uniform\\
   
\bottomrule
 \end{tabular}
\label{tab:guidelines}
\end{table*}

\vspace{2mm}
\subsubsection{Computation of Fitness Values}

In many problems, the computation of the fitness function can be extremely computationally intensive, thus motivating research in fitness approximation \cite{jin2005comprehensive}. An important practical issue is to determine ways of modeling large complex systems so that the fitness value can be computed without executing the real-world system itself (which can be impractical in many situations). 

\vspace{2mm}
\subsubsection{Setting of Termination Criteria}

Assuming the population, and the various GA control variables, have been set up properly, the GA will improve towards the optimal solution over successive generations. In light of potentially slow convergence, it is very important to for timely problem solving with GAs to devise a suitable termination criterion that defines a `good enough' solution. Defining these criteria is non-trivial and context-dependent. While this criteria depends on the number of objective functions, and the number of variables, a default value of 2.5 times the population size can be used as a reasonable maximum generation count estimate \cite{cox2005fuzzy}. 

\section{Open Issues and Future Work}
\label{sec:openissues}

GAs have become a powerful tool for solving almost any kind of an optimization problem--subject to proper modeling and encoding of chromosome. GAs are contributing in military, disaster management, spectrum utilization, and network optimization. There are numerous challenges in devising an appropriate GA based solution including suitable definitions of population size and evaluation function (which is typically non-trivial to define). In this section, we provide the recommendation on scope of future research on GAs in the field of networking. 

\subsection{Theoretical Results for GAs}

A lot of effort has focused on providing theoretical understanding of why and how GAs work but rigorous results have been slow in coming. Some of the results, such as ``No Free Lunch'', have sobering implications. Simply stated, no free lunch implies that all search algorithms perform equally when averaged over all problems. This further entails that if we are comparing GAs with other search heuristics (such as hill climbing or simulation annealing), even if GA performs better for some class of problems, the other algorithms will necessarily perform better on some problems outside this class. GAs are also population-based heuristics; even as the entire population is improved by GAs, the same cannot be guaranteed for an individual existing within that population. There is a need for enhanced theoretical understanding of various facets of GAs to lay a strong foundation upon which the GA user's trust can be safely placed. 

\subsection{How to Exploit Problem Specific Knowledge?}

Techniques like GA and neural networks are often resorted to as black-box solutions to complex problems where the mechanisms to be used for problem solving are not well understood. Even without problem specific knowledge, GAs can allow insights to develop thorough identification of similarities between highly fit individuals \cite{goldberg1988genetic}. GAs exploit this information by combining highly fit similarities to explore beyond the currently known best solutions. When the subject knowledge is available, in the form of good understanding of the underlying problem and plausible candidate solutions, it is desirable to exploit this information to expedite GA's progression towards a solution. One way of embedding problem-specific information in the GA system is by seeing GAs with good structures and populations thereby accelerating search and learning progress. However, the use of problem specific knowledge to generate specialized heuristics or operators can entail some loss of generality of the GA system \cite{goldberg1988genetic}. The question of how to embed wireless-specific subject knowledge into the GA frameworks is very much an open research issue requiring further exploration. 

\subsection{Efficient Multi-Objective and Distributed GAs}

The development of practical algorithms for wireless networks often requires the use of multi-objective optimization performed in a distributed fashion. Currently, the most popular approaches in wireless networks depend on centralized GA \cite{lai2007effective}: e.g., in mobile cellular networks, the  GA runs on the BS and helps decide the cluster size and nodes deployment. In wireless networks, there is often a need to distribute the computational load to various network nodes. This can be done by a locally run GA on each node combined with local search techniques for robustness. There is an ongoing research on dynamic cluster design based on multiple metrics. We can also use distributed GA to provide maximum disjoint covers to enhance the coverage area and network life. There is room for improving the throughput by extending the GAs from single cell to multi-cell optimization, with inter-channel noise cancellation \cite{ayyadurai2011multihop}.


\subsection{GAs, Complexity Theory, and Wireless Networks}

It has been stated in \cite{page2010diversity} that a system can be called a `complex system' if the agents that constitute this system incorporate: \textit{(i)} diversity, \textit{(ii)} connectivity, \textit{(iii)} interdependence, and \textit{(iv)} adaptation. Modern wireless networks, by these definitions, define a complex system, and are thus amenable to analysis, modeling, and optimization techniques from complexity theory. Already, there has been work in applying complexity science to networks \cite{mitchell2006complex} \cite{strogatz2001exploring}, but more research is needed, particularly for the specific case of wireless networks.

\subsection{Using Big Data Techniques For Scaling GAs}

Big data techniques have the ability to scalably and reliably process massive amounts of data using a cluster of commodity servers \cite{qadir2014acmfit}. Big data processing techniques, such as Google's MapReduce and other frameworks of its ilk, hold a lot of promise for processing large problems in quick time. This is done by horizontal scaling, or  `scaling out', through which more and more computers are employed  to attack a given problem in parallel. There is potentially a lot of scope in employing these big data processing techniques to accelerate the convergence of GAs. A steady trickle of research has started emerging that is proposing the use of big data tools (such as the open-source Apache Hadoop platform which is built on Google's MapReduce paradigm) to scale up GAs \cite{verma2009scaling} \cite{llora2010huge}. More research needs to be done to determine how to best utilize big data tools and techniques for improving GA performance.

\vspace{10pt}
\section{Conclusions}
\label{sec:conclusions}

We have provided a survey of the applications of genetic algorithms (GAs) in wireless networking. In addition to providing a self-contained introduction to common models and configurations of GAs, we have also provided a detailed survey of applications of GAs in wireless networking. We have categorized the applications of GAs in wireless networking both according to the wireless networking configuration, and according to the different kinds of GA techniques. We have also highlighted pitfalls and challenges in successfully implementing GAs in wireless networks. Finally, we have highlighted a number of open issues and have identified potential directions for future work.

\bibliographystyle{ieeetr}
\bibliography{GA}


\end{document}